\documentclass[12pt]{article}  %fuer beidseitige Version,
\usepackage[a4paper,textwidth=490.0pt,textheight=703.1pt]{geometry}
\usepackage{textcomp}
\usepackage{dutchcal}
\usepackage{hyperref}
\usepackage{slashed}
\usepackage{graphicx}
\usepackage{amsmath}
\usepackage{amssymb}
\usepackage{color}                                                  
\usepackage{float}
\usepackage[rflt]{floatflt}
\usepackage{cite}
\usepackage{mathtools}
\usepackage{dsfont}
\usepackage{subfig}

\newcommand{\HA}{{\rm H}}

\numberwithin{equation}{section}

\usepackage{array}

\usepackage[english]{babel}
\usepackage[latin1]{inputenc}
\usepackage[T1]{fontenc}
\usepackage{ae}

\usepackage{url}

\usepackage{amsmath, amsthm, amssymb}

\newtheorem{thm}{Theorem}[section]

\usepackage{array}

\usepackage[english]{babel}
\usepackage[latin1]{inputenc}
\usepackage[T1]{fontenc}
\usepackage{ae}

\newtheorem{definition}[thm]{Definition}

\newcommand\e{{\rm e}}
\newcommand{\Li}{{\rm Li}}
\newcommand{\Mvec}{{\rm \bf M}}

\catcode`,\active

\catcode`\,12

\usepackage{rotating}
\newcommand{\shuffle}{\, \raisebox{1.2ex}[0mm][0mm]{\rotatebox{270}{$\exists$}} \,}
\usepackage{graphicx}
\newcounter{mmacnt}
\def\restartmma{\setcounter{mmacnt}{0}}
\restartmma \catcode`|=\active
\def|#1|{\mathrm{#1}}
\catcode`|=12
\newenvironment{mma}{
 \par\smallskip
 \catcode`|=\active
 \parskip=0pt\parindent=0pt % locally
 \small
 \def\In##1\\{%
   \def\linebreak{\hfill\break\null\qquad}%
   \refstepcounter{mmacnt}
   \hangindent=2.5em\hangafter=0
   \leavevmode
   \llap{\tiny\sffamily In[\arabic{mmacnt}]:=\kern.5em}%
   \mathversion{bold}\footnotesize$\displaystyle##1$\normalsize
   \mathversion{normal}\par
 }%
 \def\Print##1\\{%
   \def\linebreak{\hfill\break}%
   \hangindent=2.5em\hangafter=0
   \leavevmode ##1\par}%
 \def\Out##1\\{%
   \def\linebreak{$\hfill\break\null\hfill$}%
   \kern\abovedisplayskip\par
   \hangindent=2.5em\hangafter=0
   \leavevmode
   \llap{\tiny\sffamily Out[\arabic{mmacnt}]=\kern.5em}
   \footnotesize$\displaystyle##1$\normalsize\hfill\null\par
   \kern\belowdisplayskip
 }%
 \def\Warning##1##2\\{%
   \def\linebreak{\hfill\break}%
   \hangindent=2.5em\hangafter=0
   \leavevmode
   {\scriptsize##1 : ##2}\par}%
}{%
 \par\smallskip
}

\allowdisplaybreaks[4]

\usepackage{color}

\newenvironment{fshaded}{%
\MakeFramed {\FrameRestore}
}%
{\endMakeFramed}

\usepackage{tikz}
\usetikzlibrary{matrix}

\allowdisplaybreaks[4]

\begin{document}
\setlength{\baselineskip}{0.515cm}
\sloppy
\thispagestyle{empty}
\tikzset{
	graviton/.style={decorate,line width=0.25mm, decoration={snake,amplitude=.5mm, segment length=2mm}},
	massive/.style={postaction={decorate},
		line width=0.4mm,
	},
}
 
\makeatletter
\def\simgt{\mathrel{\lower2.5pt\vbox{\lineskip=0pt\baselineskip=0pt
           \hbox{$>$}\hbox{$\sim$}}}}
\def\simlt{\mathrel{\lower2.5pt\vbox{\lineskip=0pt\baselineskip=0pt
           \hbox{$<$}\hbox{$\sim$}}}}
\makeatother
           
\def\draftnote#1{{\textcolor{red}{\it #1}}}

\def\FT#1{{\color{magenta} [FT: #1]}}
\def\DK#1{{\color{blue} [DK: #1]}}
\def\TS#1{{\color{cyan} [TS: #1]}}

\def\fig#1{Fig.~\ref{#1}}

\def\eqn#1{Eq.~\eqref{#1}}
\def\eqns#1.#2{Eqs.~\eqref{#1} and~\eqref{#2}}
\def\spa#1.#2{\left\langle#1\,#2\right\rangle}
\def\spb#1.#2{\left[#1\,#2\right]}
\def\sand#1.#2.#3{%
\left\langle#1{\vphantom1}\right|{#2}\left|#3\right]}%
\def\sandmp#1.#2.#3{%
\left\langle#1{\vphantom1}\right|{#2}\left|#3\right]}%
\def\sandpm#1.#2.#3{%
\left[#1{\vphantom1}\right|{#2}\left|#3\right\rangle}%
\def\sandmm#1.#2.#3{%
\left\langle#1{\vphantom1}\right|{#2}\left|#3\right\rangle}%
\def\sandpp#1.#2.#3{%
\left[#1{\vphantom1}\right|{#2}\left|#3\right]}%

\def\pp{\sigma}
\def\LL{\mathcal{L}}
\def\phis{\phi_{s}}
\def\CC{C_{2}}
\def\PLS{\mathbb{S}}
\def\Es{\mathcal{E}}
\def\gS{\mathsf{S}}
\def\oS{\mathsf{S}}
\def\sS{\mathsf{S}}
\def\order{O}
\def\op{\mathcal{O}}
\def\opK{\mathcal{K}}

\def\hdelta{{\hat\delta}}
\def\opa{{\hat a}}
\def\opT{{\mathbb{T}}}

%%% Hamiltonian
\def\opH{\mathcal{H}}
\def\opp{\bm p}
\def\opr{\bm r}
\def\opL{\left(\opr \times \opp\right)}
\newcommand{\spinStr}[1]{\Sigma_{#1}}
\def\clS{\textbf{S}}
\def\clK{\textbf{K}}
\def\KxS{X}
\def\clKxS{\textbf{\KxS}}
\def\eftSigma{\sigma}

\def\doe{\partial}
\def\bs{\boldsymbol}
\def\mc{\mathcal}
\def\clp{\bm p}
\def\clpb{\bar{\bm p}}
\def\clq{\bm q}

\newcommand{\sym}[1]{\{#1\}}

\def\nn{\nonumber}
\def\vs{\vskip 0cm }

\newcommand*\pFqskip{8mu}
\catcode`,\active
\newcommand*\pFq{\begingroup
        \catcode`\,\active
        \def ,{\mskip\pFqskip\relax}%
        \dopFq
}
\catcode`\,12
\def\dopFq#1#2#3#4#5{%
        {}_{#1}F_{#2}\biggl[\genfrac..{0pt}{}{#3}{#4};#5\biggr]%
        \endgroup
}

\newcommand{\MeV}{\rm MeV}
\newcommand{\GeV}{\rm GeV}
\newcommand{\be}{\begin{equation}}
\newcommand{\ee}{\end{equation}}
\newcommand{\eq}[2]{\be\begin{aligned}#1 \label{#2}\end{aligned}\ee}

\newcommand{\Fig}[1]{Fig.~\ref{#1}}
\newcommand{\Eq}[1]{Eq.~\eqref{#1}}
\newcommand{\Eqs}[2]{Eqs.~\eqref{#1} and \eqref{#2}}
\newcommand{\Sec}[1]{Sec.~\ref{#1}}
\newcommand{\Secs}[2]{Secs.~\ref{#1} and \ref{#2}}
\newcommand{\App}[1]{App.~\ref{#1}}
\newcommand{\vev}[1]{\langle #1 \rangle}
\newcommand{\bra}[1]{\langle #1 |}
\newcommand{\ket}[1]{| #1 \rangle}

\newcommand{\sslash}[1]{\ensuremath\raisebox{-0.00cm}{{\small\slash}}\hspace{-0.21cm}#1\/}
\newcommand{\dd}[1]{\frac{\partial}{\partial #1}}

\newcommand{\OurOrder}{  {\cal{O}}(G^3)  }

\newcommand{\ECM}{E_{\rm CM}}
\newcommand{\pCM}{\bm{p}_{\rm CM}}

\newcommand{\mbf}[1]{\mathbf{#1}}

\newcommand{\AEFT}{A_{\rm EFT}}

\newcommand{\MEFT}{M_{\rm EFT}}

\newcommand{\II}{{\cal I}}

\newcommand{\E}{{\rm E}}
\newcommand{\K}{{\rm K}}

\newcommand{\gfkt}[3]{\ell_{#1,#2}^{(#3)}}
\newcommand{\pot}{\rm pot}

\renewcommand{\imath}{\mathrm{i}}

\def\topbotatom#1{\hbox{\hbox to 0pt{$#1\bot$\hss}$#1\top$}} \newcommand*{\topbot}{\mathrel{\mathchoice{\topbotatom\displaystyle} {\topbotatom\textstyle} {\topbotatom\scriptstyle} {\topbotatom\scriptscriptstyle}}}

\newcommand{\tabeq}[2]{ \parbox{#1}{  \be\begin{aligned}#2 \end{aligned} \nonumber \ee }}

%%%\title{{\sf \footnotesize 
\begin{flushleft}
DESY 26--100 \hfill     % {\tt arXiv:2606.xxxxx [hep-th]}
\\ 
RISC Report number 26--11  \hfill %\today 
\\
\end{flushleft}

\vspace*{2cm}
\begin{center}
{\Large \bf \boldmath The $q$-extension of iterated integrals and nested sums}

\vspace*{2mm}
{\Large \bf in quantum field theory}

\vspace*{30mm}
\normalsize
J. Bl\"umlein$^{a,b}$, A.M.~Gavrilik$^c$,  O.~Mykhailiv$^c$ and C.~Schneider$^d$

\vspace*{5mm}
{\it $^a$Deutsches Elektronen-Synchrotron DESY, Platanenallee 6, 15738 Zeuthen, Germany}

\vspace*{2mm}
{\it $^b$Institut f\"ur Theoretische Physik III, IV, TU Dortmund, \\ Otto-Hahn 
Stra\ss{}e 4, 44227 Dortmund, Germany}

\vspace*{2mm}
{\it $^c$
Bogolyubov Institute for Theoretical Physics, 14-B Metrolohichna str., Kyiv, 03143, Ukraine}

\vspace*{2mm}
{\it $^d$
Johannes Kepler University Linz, Research Institute for Symbolic
Computation (RISC),\\ Altenberger Stra\ss{}e 69, A-4040, Linz, Austria}
\end{center}

\vspace*{3cm}
\begin{abstract}
  \noindent
  Analytic calculations of zero- and single-scale quantities in perturbative quantum field
  theory result into special numbers and functions, the first of which  have been revealed 
  during the last decades. These are generalizations of the polylogarithm in form of 
  Kummer-Poincar\'e iterative integrals over special alphabets and extensions thereof.
  With growing order in the coupling constant, the polylogarithms, Nielsen integrals, the 
  iterated integrals over linear denominator terms, cyclotomic letters, letters induced by 
  quadratic forms, square-root valued letters, and more general functions contribute. 
  For the nested sums we consider nested harmonic sums, generalized harmonic sums,
  nested sums implied by quadratic forms, cyclotomic harmonic sums, and nested sums
  containing central binomials. We construct the $q$-extensions of these special functions 
  and of the nested sums, which are associated to them by the series expansion at $x=0$, and 
  their Mellin transform in the $q$-free case. These functions are expected to 
  play a role in perturbative calculations in the case of $q$-deformed commutation relations. 
  For the simpler function spaces closed form solutions are presented. For more involved
  alphabets we present the algorithmic steps leading to the $q$-extension for the individual 
  cases. We also derive the determining differential and difference equations 
  of these higher transcendental functions. The $q$-extended special functions are
  quite different form the corresponding $\mu$-extended functions.
\end{abstract}

%\maketitle

\newpage
%----------------------------------------------------------------------------------------------------------------
\section{Introduction}  
\label{sec:1}
%----------------------------------------------------------------------------------------------------------------

\vspace*{1mm}
\noindent
The $q$-extension of mathematical functions, i.e. the association of the so-called 
basic-function to a given function, has a very long history, see Ref.~\cite{ERNST}.
$q$-analogues have been calculated for many special functions, as described in the 
surveys Refs.~\cite{HEINE1,HEINE2,BAILEY,SLATER,EXTON,GASRHA,KOORNWINDER,PWZ,ANDREWS,KOEKOEK,
KACH,NIST,ISMAIL,KOEPF,ISMAIL1,JOHNSON}. In particular, the $q$-extension of the 
generalized hypergeometric functions and their 
generalizations, like the Appell-functions,
Lauricella-functions and others, Ref.~\cite{BAILEY,SLATER,EXTON,Blumlein:2021hbq,
Passarino:2024ugq}, play a central role, cf. Refs.~\cite{6,77,78}, since these functions
cover large classes of special functions. It applies to different type orthgonal 
polynomials \cite{KOORNWINDER,KOEKOEK}.\footnote{It concerns the $q$-extensions of the 
Al-Salam-Chihara polynomials \cite{31}, Jacobi polynomials \cite{32,33,34,35,36,37}, 
Legendre polynomials \cite{38,39,40,41}, Gegenbauer polynomials \cite{42,43,44}, 
Laguerre polynomials \cite{45}, Koornwinder polynomials \cite{46}, Hermite polynomials 
\cite{47,48}, Lommel polynomials \cite{49,50}, Krawtchouk polynomials \cite{51,52,53}, 
Chebyshev polynomials \cite{56a}, Meixner polynomials \cite{54}, Charlier polynomials 
\cite{55}, Wilson polynomials \cite{56}, Hahn polynomials \cite{6}, the $q$-binomial 
polynomials, called Rogers-Szeg\"o polynomials \cite{58,59,60}, and Hall-Littlewood 
symmetric functions \cite{62}, the $q$-binomial coefficient or Gaussian polynomial with 
extension to $q$-multinomial coefficients \cite{64,65}, the $q$-disk polynomials \cite{66,67}, 
and the Macdonald polynomials, related to the one-row Young diagrams \cite{68,69,70}.} 
Also the Bessel functions belong to the class of hypergeometric solutions, which are 
extended to the Jackson $q$-Bessel and Hankel functions 
\cite{JACKSON06,GASRHA,45,79,80,81,82,83,84,85}. The Al-Salam-Chihara polynomials are 
closely related to representations of the quantized universal enveloping algebra 
$U_q(su(1,1))$. The Hall-Littlewood symmetric functions \cite{62}, with extension to the 
multivariate Rogers-Szeg\"o polynomials $H_{n}$ and generalized Galois numbers, are linked 
to affine Kac-Moody algebras \cite{63}. Many of the $q$-orthogonal polynomials occupy 
distinguished positions within the Askey scheme \cite{30,KOEKOEK} and play a central role 
in harmonic analysis \cite{73} and algebraic combinatorics.

The $q$-extensions are applied to quantum mechanical systems modifying the Heisenberg
algebra by the $q$-deformation \cite{Schwenk:1992sq,Wess:1998ht} 
%---------------------------------------------------------------------------------------
\begin{eqnarray}
\hat{x} \hat{p} - q \hat{p} \hat{x} = i, 
\end{eqnarray}
%---------------------------------------------------------------------------------------
where $q$ denotes the parameter of the extension, and $\hat{x}$ and $\hat{p}$ are 
the position and the momentum operators. 

Within this framework, $q$-analogues in quantum 
mechanics, such as for the Schr\"odinger equation, conformal quantum mechanics, and 
harmonic oscillator systems were constructed and analyzed \cite{ARIK1,BIEDENHARN,MACFARLANE,
130,131,132}, as well as Lie algebra representations were given in 
Refs.~\cite{JIMBO,DRINFELD,FADDEEV3,
HAYASHI,AIZAWA}. Extensions were made to $p,q$-deformed oscillators 
\cite{Chakrabarty,Aric,G10,G11} linked to the quantum algebra $su(2)_{p,q}$ and a 
large number of versions of $q$-oscillators were considered 
\cite{ODAKA,CHATURVEDI,G8,G9,G21}, as
contained in the $p,q$-family of deformed oscillators. Furthermore, 3-, 4-, and even 
5-parametric deformed oscillators proposed in \cite{CHUNG,BOROZOV,BURBAN1} were studied 
and so-called  $k$-bonacci \cite{CHUNG4} or quasi-Fibonacci \cite{G20} structures were 
found. Deformed Heisenberg algebras were investigated in Refs.~\cite{Gavrilik:2012yj,Gavrilik:2015sfa,
G31,G21}. Also, it is worth to emphasize that deformed oscillator algebras are naturally 
considered as the respective modified `boson' algebras, cf. Refs.~ 
\cite{G32,G33,G34,G35}.

Quantum mechanical $q$-extensions were also used for the thermostatistical description 
of $q$-bosons in Refs.~\cite{Anchishkin1,Anchishkin2,Anchishkin3,Adamska,G60,8a,9a,G61,10a}.
Another important application is the use of $q$-analogues for the description 
of hadron flavor symmetries in Refs.~\cite{G40,G41,G42,G43} 
based on such 
$q$-deformed structures as quantum groups $SU_q(N_{F})$ and quantum algebras $U_q(su_N)$ in
Refs.~\cite{G40,G42,G43,G44,G45,G46,G47,G48,G49,G50,G51}. In these analyses new hadron mass 
sum rules were obtained at high precision \cite{G40,G43,G46,G47,G48,G49,G50,G51,G52,G53}.
In several applications \cite{49a,50a} the Lerch transcendent $\Phi(z, \alpha, \beta)$, 
Ref.~\cite{LERCH}, which covers a wide set of functions, occurs.
In Ref.~\cite{G61} $(p,q)$-extended polylogarithms and the associated  
$\zeta$-function emerge. Bi-basic hypergeometric functions \cite{GASRHA} occured in 
Ref.~\cite{G20}. In supersymmetric extensions \cite{52a,53a} twin-basic hypergeometric
functions \cite{GASRHA} emerge. The $q$-Gegenbauer polynomials \cite{35}
are applied in the $q$-extension
of spin-network approaches \cite{55a}. In
meson mass sum rules \cite{G40,G47}, certain $q$-polynomials, such as the 
$q$-Chebyshev polynomials \cite{56a}, emerge.

The $q$-extension in quantum field theories 
are implied by the modified commutation (or anti-commutation) relations
%---------------------------------------------------------------------------------------
\begin{eqnarray}
\label{eq:com}
a_\lambda(k) a_{\lambda'}^\dagger(k') - q a^\dagger_{\lambda'}(k') a_\lambda(k)
=  g_{\lambda \lambda'} \delta^{(3)}(k-k'), 
\end{eqnarray}
%---------------------------------------------------------------------------------------
as shown in~Ref.~\cite{Arefeva:1995rpr,WACHTER}. Here $a_\lambda(k)$ and 
$a_{\lambda'}^\dagger(k')$ are the creation and annihilation operators.
The $q$-modified quantum field theories are related to approaches based on non-commutative space time
\cite{CONNES,CONNES1,Kowalski-Glikman:2002eyl,Chaichian:1992fx} and quantum groups
\cite{MANIN,KASSEL,CHARI,KS,Toller:2003yz}. In the study of quantum groups, conformal 
field 
theory, and of integrable systems, the theory of the $q$-Virasoro algebra plays an 
important role, cf.~Refs.~\cite{95,96,97}. 
Also $q$-supersymmetric theories 
and $q$-superalgebras have been studied, cf.~Refs.~\cite{52a,107,108,109,110,111}.

The scope of the present paper is to construct the $q$-extensions of the different 
spaces of special functions occurring in perturbative  calculations in quantum field 
theories of scattering processes to higher loop order in the non-modified case.
These are classes of iterated integrals, such
as polylogarithms \cite{LEIBNIZ1,MAXIMOM,SPENCE1,Jonquiere1,LEWIN1,LEWIN2,Devoto:1983tc}, 
Nielsen integrals 
\cite{NIELSEN1,Kolbig:1983qt}, 
followed by the harmonic polylogarithms \cite{Remiddi:1999ew}. 
They are special sub-spaces of the generalized harmonic polylogarithms,
which were first introduced by E.~Kummer, 
see~Refs.~\cite{KUMMER1,KUMMER2,KUMMER3,POINCARE1,LAPPO,CHEN,GONCHAROV,Moch:2001zr,
Ablinger:2013cf}. Other extensions are the 
cyclotomic harmonic polylogarithms \cite{Ablinger:2011te} and iterated integrals 
inspired by general quadratic forms \cite{Ablinger:2021fnc}.
In these classes the denominator functions are either linear functions or 
polynomials. There is a further class of functions obeying first order factorizing 
differential equations, based on alphabets containing square-root valued letters 
\cite{Ablinger:2014bra}. 

The series expansion of these functions around $x=0$ or their Mellin transform
lead to different classes of nested sums. These are the harmonic sums 
\cite{Vermaseren:1998uu,Blumlein:1998if}, the generalized harmonic 
sums~\cite{Moch:2001zr,Ablinger:2013cf}, and sums, which additionally contain 
central binomial factors \cite{Ablinger:2014bra}. These quantities have been 
reviewed in detail in Ref.~\cite{Blumlein:2026clg}. In this paper we will  derive 
their $q$-extension. 
In a preceding paper \cite{Blumlein:2026clg} we have studied the so-called 
$\mu$-extension, 
see Refs.~\cite{JANU,Gavrilik:2010xu}
of the mentioned classes of generalized special functions.
These extensions are, however, structurally very different 
from the $q$-extensions.

Like in the case of the $\mu$-extension, the starting point is the closed-form solution 
$f[n]$
for the respective function\footnote{Also modifications of Eq.~(\ref{eq:2}) may be 
considered, see Ref.~\cite{Blumlein:2026clg}, Eqs.~(6,7). For sums starting at $n = 1$ we 
add $f[0] = 0$ symbolically for technical reasons in the calculations below.} is
%-----------------------------------------------------------------------------------------------------
\begin{eqnarray}
\label{eq:2}
F(x) = \sum_{n=0}^\infty f[n] x^n. 
\end{eqnarray}
%-----------------------------------------------------------------------------------------------------
The functions $f[n]$ have a sum-product representation in terms of nested sums. 
This function must be known in closed form for all integers $n \in \mathbb{N}$.
The 
$q$-extension of $F(x)$ is obtained from the $q$-extension of $f[n]$, $f[n;q]$,\footnote{
We also use the notation
$\langle x^n \rangle F(x;q) = f[n;q],~~~n \geq 0.$ }
%-----------------------------------------------------------------------------------------------------
\begin{eqnarray}
\label{eq:2a}
F(x;q) = \sum_{n=0}^\infty f[n;q] x^n. 
\end{eqnarray}
%-----------------------------------------------------------------------------------------------------
For $q$-extended nested sums and iterated integrals a number of algorithms were created,
like the extension of Zeilberger's algorithm \cite{ZEILBERGER1,ZEILBERGER2 } in 
Refs.~\cite{WZ1,Koornwinder3,
RIESE,BK,LE,CHM,PR,RIESE5}  and Refs.~\cite{SIG1,SIG2,RIESE4,Ocansey,Chen1}, as well as
operator methods for $q$-identities \cite{CIGLER1,CIGLER2,CIGLER3}.

Compared to the $\mu$-extension of the different function classes studied in
Ref.~\cite{Blumlein:2026clg}, the $q$-extensions define new function spaces of
higher transcendental functions. Already the simplest quantities are given by 
$q$-hypergeometric functions. The corresponding difference and differential 
equations grow with the weight of the extended quantities.
The main characteristics defining the respective $q$-extensions are the recurrence 
relations for the quantities $f[n;q]$ and the $q$-differential and $q$-shift relations 
for $F[x;q]$. The $q$-extended nested sums are defined by their recurrence relation.
If the recursion for $f[n;q]$ is known, a fast numerical evaluation of $F(x;q)$, 
Eq.~(\ref{eq:2a}), 
around $x = 0$ is possible. Series representations of $F(x;q)$ have a finite convergence 
radius. 
To obtain the $q$-extension around a different point $x_0$, a
further series expansion of $F(x)$ must be performed, ensuring that the domains
of convergence for neighboring series overlap.
Because of its similarity to Frobenius' method \cite{FROBENIUS}, which is in wide use
in quantum field-theoretic calculations, see, 
e.g.,~\cite{Grigo:2012ji,Fael:2022miw,Behring:2023rlq}, we call 
this method {\it $q$-Frobenius method}. 

The paper is organized as follows. In Section~\ref{sec:2}  we summarize basic 
definitions and operations used to construct the $q$-extensions. All nested sums 
and iterated integrals for which we are constructing the $q$-extension obey first order 
factorizing difference 
and differential equations. A summary on these quantities has been given by us 
in Ref.~\cite{Blumlein:2026clg} previously. The nested sums and iterated integrals 
are built over special alphabets of letters, $\mathfrak{S}$ or $\mathfrak{A}$. 
The $q$-extensions of the different classes of nested sums 
are derived in Section~\ref{sec:3} and the ones for the iterated integrals in 
Section~\ref{sec:4}. Here we also derive the defining equations. 
For the simple cases, such as the single harmonic sums, the polylogarithms and Nielsen integrals, 
one can derive general representations. For the more involved
alphabets we device the sequence of algorithmic steps to obtain the result in each 
individual case.
Like in the $q$-free case, only a finite
amount of nested finite sums and iterated integrals contributes in 
perturbative field-theoretic calculations up to a certain power in the coupling constant.
In Section~\ref{sec:5} we discuss the shuffle algebras for the $q$-extended 
iterated integrals and the quasi-shuffle algebras for the $q$-extended nested sums. 
Their use allows one to significantly reduce the number of contributing terms at a given weight.
Section~\ref{sec:6} contains the conclusions.
%----------------------------------------------------------------------------------------------------------------
\section{Basic Definitions}  
\label{sec:2}
%----------------------------------------------------------------------------------------------------------------

\vspace*{1mm}
\noindent
We first summarize relations given in the literature for the asymmetric 
$q$-extension\footnote{There are also other $q$-extensions, see Appendix~A of 
Ref.~\cite{Blumlein:2026clg}. The symmetric $q$-extension, which obeys the symmetry $q 
\leftrightarrow q^{-1}$ is applied in Refs.~\cite{KS,BIEDENHARN,MACFARLANE}, two-variable 
extensions were used in Refs.~\cite{Chakrabarty,Aric,BURBAN3}, and multi-parameter 
extensions
in Refs.~\cite{CHUNG,BOROZOV,BURBAN1,BURBAN,Gavrilik:2012yj,Gavrilik:2015sfa}.}, 
which we use in the following, see 
Refs.~\cite{ERNST,HEINE1,HEINE2,BAILEY,SLATER,EXTON,GASRHA,KOORNWINDER,ANDREWS,KACH,NIST}.
The $q$-extension of an integer number $n$ is given by
%----------------------------------------------------------------------------------------------------------------
\begin{eqnarray}
\label{eq:BAS1}
\{n\}_q = \frac{1-q^n}{1-q} = \sum_{k=0}^{n-1} q^k,~~~\text{with}~~q \in [0,1]. 
\end{eqnarray}
%----------------------------------------------------------------------------------------------------------------
The variables $n$ occurring in powers
%----------------------------------------------------------------------------------------------------------------
\begin{eqnarray}
c^n,~~c \in \mathbb{C}, 
\end{eqnarray}
%----------------------------------------------------------------------------------------------------------------
are not $q$-extended.

The  Jackson $q$-derivative is defined by \cite{JACKSON1,JACKSON2,THOMAE1,THOMAE2}
%---------------------------------------------------------------------------------------------$
\begin{eqnarray}
\label{eq:diffq}
{\cal D}_x^{q} f[x] =  \frac{f(x) - f(qx)}{(1-q)x}, 
\end{eqnarray}
%---------------------------------------------------------------------------------------------$
with
%---------------------------------------------------------------------------------------------$
\begin{eqnarray}
{\cal D}_x^{q} [x^\alpha] =  \frac{(1-q^\alpha)}{(1-q)} x^{\alpha-1},~~\alpha \in \mathbb{R}, 
\end{eqnarray}
%---------------------------------------------------------------------------------------------$
for the power function. The operator ${\cal D}_x^{q}$ becomes only a differential 
operator in the limit $q \rightarrow 1^-$. Otherwise it is a difference operator, by 
which 
the $q$-differential (\ref{eq:qdi}) and $q$-shift relations (\ref{eq:qsh}), derived for 
the iterated integrals below, are related.

The $q$-extended Pochhammer symbol \cite{POCHHAMMER} reads, cf.~Ref.~\cite{CIGLER2},
%---------------------------------------------------------------------------------------------$
\begin{eqnarray}
(a; q)_n &=& \prod_{k=0}^{n-1} (1-a q^k),~~~(a; q)_0 = 1,   \\
(q; q)_n &=& \prod_{k=1}^{n} (1- q^k),
~~~(q^2; q)_n = \prod_{k=2}^{n+1} (1- q^k). 
\end{eqnarray}
%---------------------------------------------------------------------------------------------$
The {\it rising} Pochhammer symbol $(x)_n$
can be written in terms of the Stirling numbers of the 
1st kind \cite{STIRLING}, respectively for their $q$-extension,
%---------------------------------------------------------------------------------------------$
\begin{eqnarray}
(x)_n   &=& \sum_{k=0}^n s(n,k)   (-1)^{n-k} x^k,  \\
(x;q)_n &=& \sum_{k=0}^n s(n,k;q) (-1)^{n-k} x^k, 
\end{eqnarray}
%---------------------------------------------------------------------------------------------$
for which the recursions
%---------------------------------------------------------------------------------------------$
\begin{eqnarray}
s(n+1,k)     &=& s(n,k-1)   - n   ~s(n,k),  \\ &&
s(0,0) = 1,~~s(n,0) = 0~~\text{for}~~n > 0,~~s(0,k) = 0~~\text{for}~~k > 0, \nonumber 
\\
s(n+1,k;q)   &=& s(n,k-1;q) - \{n\}_q ~s(n,k;q), \\ &&
s(0,0;q) = 1,~~s(n,0;q) = 0~~\text{for}~~n > 0,~~s(0,k;q) = 0~~\text{for}~~k > 0, 
\nonumber
\end{eqnarray}
%---------------------------------------------------------------------------------------------$
hold.

The $q$-extended factorial, the Gaussian $q$- and central binomial  \cite{GAUSS3}
are given by 
%---------------------------------------------------------------------------------------------$
\begin{eqnarray}
\{n!\}_q &=& \prod_{k=1}^n \{k\}_q = \Gamma_q(n+1),    \\
\binom{n}{m}_q &=& \frac{(q)_n}{(q)_m (q)_{n-m}} = \prod_{i=0}^{m-1} 
\frac{1-q^{n-1}}{1-q^{i+1}},   \\
\binom{2n}{n}_q &=& \frac{\{2n!\}_q}{\{n!\}_q^2}. 
\end{eqnarray}
%---------------------------------------------------------------------------------------------$
Due to
%---------------------------------------------------------------------------------------------$
\begin{eqnarray}
\binom{n+1}{k}_q = q^k \binom{n}{k}_q + \binom{n}{k-1}_q,~~\binom{n}{1}_q 
= \frac{1-q^n}{1-q},~
\binom{n}{2}_q = \frac{(1-q^n)(1-q^{n-1})}{(1-q)(1-q^2)} 
\end{eqnarray}
%---------------------------------------------------------------------------------------------$
the $q$-binomial is a polynomial, since the initial values are polynomials in $q$, because for each $n$ 
$(1-q^n)$
contains the factor $(1-q)$ and for each even $n$ the factor $(1-q^2)$.

For real argument $z$, 
%---------------------------------------------------------------------------------------------$
\begin{eqnarray}
\Gamma(z+1) = z \Gamma(z),~~~~ z \notin \{\mathbb{Z}, z <1 \},
\end{eqnarray}
%---------------------------------------------------------------------------------------------$
the function $\Gamma_q$ is defined by 
%---------------------------------------------------------------------------------------------$
\begin{eqnarray}
\Gamma_q(z+1) = \frac{1-q^z}{1-q} \Gamma_q(z),~~~~\Gamma_q(1) = 1. 
\end{eqnarray}
%---------------------------------------------------------------------------------------------$
Related to it, the digamma function is
%---------------------------------------------------------------------------------------------$
\begin{eqnarray}
\psi(z)   &=& \frac{d}{dz} \ln[\Gamma(z)], \\
\psi(z+1) &=& \frac{1}{z} + \psi(z), ~~~~ z \notin \{\mathbb{Z}, z <1 \}.
\end{eqnarray}
%---------------------------------------------------------------------------------------------$
The functions $\psi_q^{(n)}, n \geq 0,$ are given by
%---------------------------------------------------------------------------------------------$
\begin{eqnarray}
\label{qdig}
\psi_q(z) &=& \frac{d}{dz} \ln[\Gamma_q(z)] = -\ln(1-q) + \ln(q) \sum_{n=0}^\infty 
\frac{q^{n+z}}{1 - q^{n+z}},    \\
\psi_q^{(n)}(z) &=& \frac{d^n}{dz^n} \psi_q(z). 
\end{eqnarray}
%---------------------------------------------------------------------------------------------$
The last sum in (\ref{qdig}) is a modified Lambert series ${\rm L}(q)$, 
\cite{LAMBERT1,KNOPP},
%---------------------------------------------------------------------------------------------$
\begin{eqnarray}
{\rm L}(q) = \sum_{n=1}^\infty \frac{q^n}{1-q^n} = \frac{1}{\ln(q)} [\psi_q(1) + 
\ln(1-q)]. 
\end{eqnarray}
%---------------------------------------------------------------------------------------------$
The asymptotic expansion of $\psi_q(z)$ in the limit $z \rightarrow \infty$ is 
given by \cite{MOAK}
%---------------------------------------------------------------------------------------------$
\begin{eqnarray}
\psi_q(z) &\propto& \ln\left(
              \frac{1-q^z}{1-q}\right) 
              + \frac{\ln(q)}{2(q^{-z}-1)}
- \sum_{k=1}^\infty 
\frac{B_{2k}}{(2k)!}
\left(\frac{\ln(q)}{(1-q)^z}\right)^{2k} 
q^z P_{2k-2}(q^z),\\
P_n(x) &=& x(1-x) \frac{d}{dx} P_{n-1}(x)+ (n x +1) P_{n-1}(x),~~~P_0 = 1,~~n \geq 
1,~~~P_n(1)=(n+1)!, 
\nonumber\\
\end{eqnarray}
%---------------------------------------------------------------------------------------------$
with $B_{2k}$ the usual Bernoulli numbers.
The first polynomials  are
%---------------------------------------------------------------------------------------------$
\begin{eqnarray}
P_1(x) &=& 1 + x,   \\
P_2(x) &=& 1 + 4 x + x^2,  \\
P_3(x) &=& 1 + 11 x + 11 x^2 + x^3,~~{\rm etc.} 
\end{eqnarray}
%---------------------------------------------------------------------------------------------$
The $q$-polygamma functions $\psi_q^{(n)}(z)$ can be used to define the single $q$-harmonic sums, as 
the polygamma functions are also the analytic continuations of the single harmonic sums,
cf.~\cite{Blumlein:1998if}, after the proper treatment of factors of $(-1)^n$, in terms 
of a suitable Mellin transform, see Ref.~\cite{Blumlein:1998if} and Section~\ref{sec:3}.

Heine \cite{HEINE1,HEINE2} introduced the $q$-generalization of the hypergeometric 
series. The $q$-extended generalized hypergeometric series are given by
%----------------------------------------------------------------------------------------------------------------
\begin{eqnarray}
_{s+1}\phi_{s} \left( 
  \begin{matrix} 
    a_1, a_2, \dots, a_{s+1} \\ 
    b_1, b_2, \dots, b_s 
  \end{matrix} 
  ; q, z 
\right) 
&=& \sum_{n=0}^{\infty} 
\frac{(a_1; q)_n (a_2; q)_n \cdots (a_{s+1}; q)_n}
{(q; q)_n (b_1; q)_n \cdots (b_s; q)_n} z^n. 
\end{eqnarray}
%----------------------------------------------------------------------------------------------------------------
%%with
%----------------------------------------------------------------------------------------------------------------
%%\begin{eqnarray}
%%(a_1, a_2, \dots, a_{s+1})_n &=& (a_1; q)_n (a_2; q)_n \cdots (a_{s+1}; q)_n .
%%\end{eqnarray}
%----------------------------------------------------------------------------------------------------------------
The $q$-extended generalized hypergeometric series provide a flexible 
framework.\footnote{They allow also to represent the elliptic polylogarithms, 
cf.~Ref.~\cite{Passarino:2016zcd}.} which can be used for some special cases 
in the following. These functions were studied in detail 
in Refs.~\cite{BAILEY,SLATER,EXTON,GASRHA,ANDREWS,NIST,KOELINK}.

The $q$-differential equation of the $q$-hypergeometric function $u(x)~=~_{2}\phi_{1}$
is given by \cite{KOELINK}
%----------------------------------------------------------------------------------------------------------------
\begin{eqnarray}
&& 
 \Biggl\{
x(b_1 - a_1 a_2 q x) ({\cal D}_x^{q})^2
+ \left[
\frac{1-b_1}{1-q} + \frac{(1-a_1)(1-a_2) - (1 - a_1 a_2 q)}
{1-q} \right] {\cal D}_x^{q} - \frac{(1-a_1)(1-a_2)}{(1-q)^2} \Biggr\} 
\nonumber\\ 
&& \times
_{2}\phi_{1} \left(
  \begin{matrix}
    a_1, a_2 \\
    b_1
  \end{matrix}  
  ; q, u
\right) = 0. 
\end{eqnarray}
%----------------------------------------------------------------------------------------------------------------
An equivalent representation is \cite{KOELINK}
%----------------------------------------------------------------------------------------------------------------//
\begin{eqnarray}
\label{eq:sh0}
&&    [b_1-{a_1} {a_2}
   q x] u(q^2 x)
+
   [q ({a_1}
   {a_2} (q+1)
   (x-1)+{a_1}+{a_2}-1)-{b_1}] u(q x) 
\nonumber\\ &&
-[q (x-1) 
   ({a_1} ({a_2}
   q+{a_2}-1)-{a_2}+1] u(x)
 = 0, 
\end{eqnarray}
%----------------------------------------------------------------------------------------------------------------
which is also called $q$-shift relation.

For the $q$-extended nested sums we will derive difference equations
%----------------------------------------------------------------------------------------------------------------//
\begin{eqnarray}
\label{eq:qrec}
\sum_{k=0}^M s_k(n;q) f[n+k;q]  = 0, 
\end{eqnarray}
%----------------------------------------------------------------------------------------------------------------
where $s_k(n;q)$ are polynomials of $q$-exponential terms $q^{l n + m}$. 

The $q$-extended iterated integrals have three representations.\\
{\it i)}~~~~the recurrence of the $q$-extension of $f[n]$, $f[n;q]$, given by 
Eq.~(\ref{eq:qrec}). \\
{\it ii)}~~~the $q$-shift difference equation is given by
%----------------------------------------------------------------------------------------------------------------//
\begin{eqnarray}
\label{eq:qsh}
\sum_{k=0}^M r_k(x,q) F[x q^k]  = 0, 
\end{eqnarray}
%----------------------------------------------------------------------------------------------------------------
where $r_k(x,q)$ are polynomials and for $F[x q^k]$ only the main argument is displayed,
like in  Eq.~(\ref{eq:sh0}).\\
{\it iii)}~~the $q$-differential equation reads
%----------------------------------------------------------------------------------------------------------------//
\begin{eqnarray}
\label{eq:qdi}
\left[\sum_{k=0}^M p_k(x,q) ({\cal D}_x^{q})^k\right] F[x;q]  = 0. 
\end{eqnarray}
%----------------------------------------------------------------------------------------------------------------
All three relations can be transformed into each other. In the case of $q$-shift 
difference and $q$-differential equations one may use Eq.~(\ref{eq:diffq}).
In the limit $q \rightarrow 1^-$ one obtains the $q$-free case for all functions studied 
in the following.
%----------------------------------------------------------------------------------------------------------------
\section{\boldmath The $q$-extensions of nested sums}  
\label{sec:3}
%----------------------------------------------------------------------------------------------------------------

\vspace*{1mm} 
\noindent 
We consider the alphabet
%-----------------------------------------------------------------------------------------------------
\begin{eqnarray}
\mathfrak{S} = \left\{s_1(k), \ldots, s_m(k)\right\},~~~~k \in \mathbb{N}, 
\end{eqnarray}
%-----------------------------------------------------------------------------------------------------
where, $s_l(k)$ are sum-product structures. The nested finite sums are defined by 
%-----------------------------------------------------------------------------------------------------
\begin{eqnarray}
S_{a,b_1,...,b_m}(N) = \sum_{k=1}^N s_a(k) S_{b_1,...,b_m}(k), 
\end{eqnarray}
%-----------------------------------------------------------------------------------------------------
where the letters $a, b_i$ label elements out of $\mathfrak{S}$.

The following sum-types play a central role both for the representation of the 
Mellin transform Eq.~(\ref{eq:MEL}) and in the analytic series expansion of the functions 
$f(x)$ 
around $x = 0$,
%-----------------------------------------------------------------------------------------------------
\begin{eqnarray}
\label{eq:MEL}
\Mvec[f(x)](N) = \int_0^1 dx~x^{N-1} f(x). 
\end{eqnarray}
%-----------------------------------------------------------------------------------------------------

\vspace*{2mm}
\noindent
{\sf i)~Harmonic sums.}\\
The $q$-extension of the harmonic sums 
\cite{Vermaseren:1998uu,Blumlein:1998if} are given by
%---------------------------------------------------------------------------------------------$
\begin{eqnarray}
S_{b,\vec{a}}(n;q) &=& 
\sum_{k=1}^n \frac{({\rm sign}(b))^k}{\{k\}_q^{|b|}} 
S_{\vec{a}}(k;q),~~~S_\emptyset = 1, b, a_i \in \mathbb{Z} \backslash \{0\}
\\ 
&=&
\sum_{k=1}^n  ({\rm sign}(b))^k \left[\frac{1-q}{1-q^k}\right]^{|b|}
S_{\vec{a}}(k;q). 
\end{eqnarray}
%---------------------------------------------------------------------------------------------$
These sums contribute to the central functions $f[n;q]$ of the $q$-extended Nielsen 
integrals and the $q$-extended
harmonic polylogarithms. In the limit $q \rightarrow 1^-$ the usual nested sums are 
obtained.

For the single harmonic sums one obtains
%---------------------------------------------------------------------------------------------$
\begin{eqnarray}
S_1(n;q) &=& (1-q)\left\{n 
+ \frac{1}{\ln(q)} \left[\psi_q(1) - \psi_q(n+1)\right]
\right\},  
\\
S_2(n;q) &=& (1-q)^2 \left\{
n
+ \frac{1}{\ln(q)} 
  \left[\psi_q(1) - \psi_q(n+1)\right]
+\frac{1}{\ln^2(q)} 
\left[\psi_q^{(1)}(1) - \psi_q^{(1)}(n+1)\right]\right\}, 
\\
S_3(n;q) &=& (1-q)^3 \left\{
n
+ \frac{1}{\ln(q)} 
  \left[\psi_q(1) - \psi_q(n+1)\right]
+\frac{3}{2\ln^2(q)} 
\left[\psi_q^{(1)}(1) - \psi_q^{(1)}(n+1)\right]\right.
\nonumber\\ &&  \hspace*{1.5cm} \left.
+\frac{1}{2\ln^3(q)}
\left[\psi_q^{(2)}(1) - \psi_q^{(2)}(n+1)\right]\right\}, 
\end{eqnarray}
%---------------------------------------------------------------------------------------------$
etc. 

The recurrence for the single $q$-harmonic sums $S_l(N;q)$ is given by\footnote{Here 
and in the following the recurrences are written in polynomial form for the 
coefficients, i.e. leaving 
out denominator factors like $(1-q)^l$ in (\ref{eq:rec1}) etc. They are valid for $q 
< 1$.}
%---------------------------------------------------------------------------------------------$
\begin{eqnarray}
\label{eq:rec1}
&& -(1 - q^{1 + n})^l F[n;q] + \left[(1 - q^{1 + n})^l + (1 - q^{2 + n})^l\right] F[n+1;q]
- (1 - q^{2 + n})^l F[n+2;q] = 0,
\nonumber\\
&& F[1;q] = 1,~~F[2;q] = 1 + \left(\frac{1-q}{1-q^2}\right)^l 
\end{eqnarray}
%---------------------------------------------------------------------------------------------$
for $l \geq 1$. In the alternating case $l \leq -1$ one has
%---------------------------------------------------------------------------------------------$
\begin{eqnarray}
&& -(1 - q^{1 + n})^l F[n;q] + \left[(1 - q^{1 + n})^l - (1 - q^{2 + n})^l\right] F[n+1;q]
+ (1 - q^{2 + n})^l F[n+2;q] = 0,
\nonumber\\
&& F[1;q] = -1,~~F[2;q] = -1 + \left(\frac{1-q}{1-q^2}\right)^l. 
\end{eqnarray}
%---------------------------------------------------------------------------------------------$

The recurrences for individual monomials can be obtained by using the command {\tt 
GuessqRecurrence} from 
the package {\tt qFunctions.m}, Ref.~\cite{UNCU}, or {\tt QREGuess} from the package
{\tt qGeneratingFunctions.m}, Ref.~\cite{KOUTSCHAN,KAUERS1}, or by direct construction of 
repeated use of Eq.~(\ref{eq:CONSTrec}) and similar relations. We had to apply different methods to 
find the recurrences derived in the following.\footnote{
Other implementations are, e.g., {\tt Gfun}, {\tt HYPQ}, {\tt RATE}, {\tt QSERIES}, and {\tt 
GUESS}
\cite{SZ,KRATTENTHALER1,KRATTENTHALER2,GARVAN,RUBEY}.} The number 
of expansion 
coefficients $f[n;q]$ for the expansions around $x=0$ needed to determine the recurrences
in the examples given below varies from $N = 15$ to $80$, depending on the complexity
of the problem. Related to this, the maximal order and degree in the search 
for the recurrences take values of up to $O(10)$. For the iterated integrals we used the 
procedures {\tt QRE2DE} and {\tt QRE2SE} \cite{KOUTSCHAN} 
to find the $q$-differential and 
$q$-shift relations.
%-------------------------------------------------------------------------------------

Examples for $q$-extended nested harmonic sums are  
%-----------------------------------------------------------------------------------------------------
\begin{eqnarray}
\label{eq:S21}
S_{2,1}(n;q) &=& \sum_{k=1}^n \left[\frac{1-q}{1-q^k}\right]^2
\sum_{l=1}^k  \frac{1-q}{1-q^l},  \\
\label{eq:S3m2}
S_{3,-2}(n;q) &=& \sum_{k=1}^n \left[\frac{1-q}{1-q^k}\right]^3
\sum_{l=1}^k (-1)^l \left[\frac{1-q}{1-q^l}\right]^2. 
\end{eqnarray}
%-----------------------------------------------------------------------------------------------------
The recurrence of $S_{2,1}(n;q)$ reads
%-----------------------------------------------------------------------------------------------------
\begin{eqnarray}
&&     
   \left(q^{n+1}-1\right)^2
   \left(q^{n+2}-1\right) F[n;q]
-
   \left(q^{n+2}-1\right)
   \left(-\left((q+1) (q+2)
   q^{n+1}\right)+\left(q^3+q^2+1
   \right) q^{2 n+2} \right.
\nonumber\\ && \left.
+3\right) F[n+1;q]
+
   \left(q^{n+2} \left(-\left((q
   (3 q+2)+4)
   q^{n+2}\right)+\left(q^3+q+1\right) q^{2 n+4}+4
   q+5\right)-3\right) \nonumber\\ 
&&
\times
F[n+2;q]
- 
   \left(q^{n+3}-1\right)^3 F[n+3;q]
= 0,
\end{eqnarray}
%-----------------------------------------------------------------------------------------------------
and for $S_{3,-2}(n;q)$
%-----------------------------------------------------------------------------------------------------
\begin{eqnarray}
&& 
\left(q^{n+1}-1\right)^3 \left(q^{n+2}-1\right)^2
   F[n;q]
+\left(q^{n+2}-1\right)^2 
\nonumber\\ &&
\times \left(\left(q
   \left(\left(3-(q-1) q^2
   (q+3)\right)
   q^n+\left(q^5-q^3-1\right) 
   q^{2 n+1}+2
   q-2\right)-3\right)
   q^{n+1}+1\right) 
F[n+1;q]
\nonumber\\ &&
+\left(\left((q (11
   q+6)-7) q^{n+2}+(9-q (q (10
   q+3)+6)) q^{2 n+4}+\left(q
   \left(5 q^3+3
   q+2\right)-5\right) q^{3
   n+6} \right.  \right.
\nonumber\\ && \left. \left.
-\left(q^5+q^2-1\right)
   q^{4 n+8}-7 q+2\right)
   q^{n+2}+1\right)
   F[n+2;q]
+\left(q^{n+3}-1\right)^5
   F[n+3;q] = 0. 
\end{eqnarray}
%-----------------------------------------------------------------------------------------------------
For the initial values of these recurrences, and analogously for those given below, we 
use the first 
values of the respective analytic expressions, like of Eqs.~(\ref{eq:S21}, \ref{eq:S3m2}).

For analytically known nested product-sum structures one can also construct the associate 
recurrence by systematic deconstruction using shift relations down to zero. As an example 
we consider the sum $S_{1,1}(n;q)$. By using 
%----------------------------------------------------------------------------------------------------- 
\begin{eqnarray} \label{eq:CONSTrec} \frac{1-q^{n+1}}{1-q} \left[S_{1,1}(n+1;q) - 
S_{1,1}(n;q)\right] = S_1(n+1;q),  \end{eqnarray} 
%----------------------------------------------------------------------------------------------------- 
a reduction is obtained, which can be iterated until the r.h.s. vanishes.
In this way one obtains the third-order linear homogeneous difference equation for 
$S_{1,1}(n;q)$,
%----------------------------------------------------------------------------------------------------- 
\begin{eqnarray} 
&& \left(q^{n+1}-1\right) \left(q^{n+2}-1\right)
   S_{1,1}(n;q)
-\left(q^{n+2}-1\right) \left(\left(q^2+q+1\right)
   q^{n+1}-3\right) S_{1,1}(n+1;q)
\nonumber\\ &&
+\left(\left(q^{n+2}-1\right)^2+\left(q^{n+3}-1\right)
   \left(q^{n+2}-1\right)+\left(q^{n+3}-1\right)^2\right)
   S_{1,1}(n+2;q)
\nonumber\\ &&
-\left(q^{n+3}-1\right)^2
   S_{1,1}(n+3;q) = 0. 
\end{eqnarray}
%-----------------------------------------------------------------------------------------------------
The common recurrence of different sum-product monomials which obey $q$-holonomic 
sequences can be obtained by using closure 
properties, cf.~Refs.~\cite{KOUTSCHAN,KAUERS2}. The command {\tt QREPlus} \cite{KOUTSCHAN} 
allows one to join two 
recurrences to a common one. We will use this procedure in some cases below.

\vspace*{2mm}
\noindent
{\sf ii)~Generalized harmonic sums.}\\
The $q$-extension of the generalized harmonic sums, 
cf.~Refs.~\cite{Moch:2001zr,Ablinger:2013cf} 
are given by
%---------------------------------------------------------------------------------------------$
\begin{eqnarray}
S_{b,\vec{a}}(\{c,\vec{d}\},n;q) &=& 
\sum_{k=1}^n \frac{c^k}{\{k\}_q^{b}} 
S_{\vec{a}}(\vec{d},k;q),~~~S_\emptyset = 1, 
b, a_i \in \mathbb{N} \backslash \{0\}, 
c, d_i \in \mathbb{C} \backslash \{0\}, 
\nonumber\\  
&=&
\sum_{k=1}^n  c^k \left[\frac{1-q}{1-q^k}\right]^{b}
S_{\vec{a}}(\vec{d},k;q). 
\end{eqnarray}
%---------------------------------------------------------------------------------------------$
These sums contribute to the central functions $f[n;q]$ of the $q$-generalized harmonic 
polylogarithms, the $q$-cyclotomic harmonic polylogarithms, and $q$-harmonic 
polylogarithms inspired by quadratic forms.

As an example we derive the recurrence for the generalized harmonic $q$-nested sum 
$S_{1,2}(-\tfrac{1}{2},2;n;q)$. It is obtained by
%---------------------------------------------------------------------------------------------$
\begin{eqnarray}
&& \left(q^{n+1}-1\right) \left(q^{n+2}-1\right)^2
   S_{1,2}(-\tfrac{1}{2}, 2; n; q)
+\left(q^{n+2}-1\right) \left(\left(q \left(q
   \left(\left(q^3+2 q-1\right)
   q^n-2\right)-3\right)+1\right) 
\right.
\nonumber\\ && \left.
\times q^{n+1} 
+2\right) 
   S_{1,2}(-\tfrac{1}{2}, 2; n+1; q)
+\left(\left(((2-5 q)
   q+6) q^{n+2}+\left(q^2 (2
   q-1)-2\right) q^{2 n+4}+4
   q \right. \right.
\nonumber\\ && \left. \left.
-7\right) q^{n+2}+1\right)
   S_{1,2}(-\tfrac{1}{2}, 2; n+2; q)
-2 \left(q^{n+3}-1\right)^3
   S_{1,2}(-\tfrac{1}{2}, 2; n+3; q) = 0. 
\end{eqnarray}
%---------------------------------------------------------------------------------------------$

\vspace*{2mm}
\noindent
{\sf iii)~Nested sums containing central binomials}\\
In these sum-structures, which contain sums of the kind {\sf ii)}, additionally factors 
of 
%---------------------------------------------------------------------------------------------$
\begin{eqnarray}
\left. \binom{2n}{n}\right|_q 
\end{eqnarray}
%---------------------------------------------------------------------------------------------$
are present, cf.~Ref.~\cite{Ablinger:2014bra}. An example will be dealt with in 
Section~\ref{sec:47}.

In summary, the $q$-extended nested finite sums obey $q$-difference equations of finite 
order 
and degree, which can be obtained by guessing methods, as, 
e.g., those of Ref.~\cite{KOUTSCHAN,UNCU}. Sums and products of these quantities form 
the expansion coefficients of the iterated integrals dealt with in Section~\ref{sec:4}.
The quasi-shuffle properties of $q$-extended sums are discussed in Section~\ref{sec:5}.
%----------------------------------------------------------------------------------------------------------------
\section{The \boldmath $q$-extended iterative integrals}  
\label{sec:4}
%----------------------------------------------------------------------------------------------------------------

\vspace*{1mm}
\noindent
The properties of the iterative integrals $F(x)$ for which we construct the $q$-extension 
in the following have been discussed in  Section~3 of Ref.~\cite{Blumlein:2026clg} 
already.
By considering their series expansion at $x=0$,
%----------------------------------------------------------------------------------------------------------------
\begin{eqnarray}
F(x) = \sum_{n=0}^\infty f[n] x^n, 
\end{eqnarray}
%----------------------------------------------------------------------------------------------------------------
the central function $f[n]$ is obtained which is needed to construct the $q$-extensions.\footnote{For deriving 
relations for functions in the $q$-free case we apply algorithms of {\tt HarmonicSums.m},
Refs.~\cite{Ablinger:2010kw,Ablinger:2013hcp}, see also Ref.~\cite{Blumlein:2026clg}.}
Here, all summation 
quantifiers and  products are $q$-extended, while powers $c^n,~~c \in \mathbb{C}$ are 
not. 
The $q$-free functions are obtained in the limit $q \rightarrow 1^-$. In some cases the 
l'Hospital rule \cite{HOSPITAL} has to be applied. 

The $q$-extensions of all iterative integrals are higher transcendental functions.
Some of them can be expressed as special $q$-extended (generalized) hypergeometric functions.
As has been outlined in Ref.~\cite{Blumlein:2026clg}, this is different in the case of 
the $\mu$-extension. There the class of special functions is preserved, except for 
the case of alphabets containing square-root valued letters.

One can derive closed form solutions for the $q$-extension of classical polylogarithms 
and Nielsen integrals. For higher classes of iterated integrals,  
because of a large amount of permutations of letters of the contributing alphabets 
$\mathfrak{A}$, we will devise methods to derive the $q$-extension for the respective 
cases. Here one has to replace the sum-product structures in $f[n]$ for all contributing 
quantities to their $q$-extension, see~Sections~\ref{sec:2} and \ref{sec:3}.

Let 
%----------------------------------------------------------------------------------------------------------------
\begin{eqnarray}
\mathfrak{A}_G =  \left\{f_1(x), \ldots f_m(x) \right\},~~~~x \in ]0,1[, 
\end{eqnarray}
%----------------------------------------------------------------------------------------------------------------
be an alphabet of functions, which are not higher transcendental. An iterated integral
is defined by 
%----------------------------------------------------------------------------------------------------------------
\begin{eqnarray}
G\left(\left\{f_b(\tau),f_{a_1}(\tau),...f_{a_k}(\tau)\right\};x\right)
= \int_0^x dy f_b(y) G\left(\left\{f_{a_1}(\tau),...f_{a_k}(\tau)\right\};y\right) = 
\sum_{n=1}^\infty f[n] x^n. 
\end{eqnarray}
%----------------------------------------------------------------------------------------------------------------
The $q$-extension of $G$ is obtained by
%----------------------------------------------------------------------------------------------------------------
\begin{eqnarray}
G\left(\left\{f_b(\tau),f_{a_1}(\tau),...f_{a_k}(\tau)\right\};x\right)_q 
= \sum_{n=1}^\infty f[n;q] x^n. 
\end{eqnarray}
%----------------------------------------------------------------------------------------------------------------
The hierarchy of iterated integrals is described by the differential equation
%----------------------------------------------------------------------------------------------------------------
\begin{eqnarray}
\frac{d}{dx}
G\left(\left\{f_b(\tau),f_{a_1}(\tau),...f_{a_k}(\tau)\right\};x\right)
= f_b(x) G\left(\left\{f_{a_1}(\tau),...f_{a_k}(\tau)\right\};x\right)  
\end{eqnarray}
%----------------------------------------------------------------------------------------------------------------
and for their $q$-extension by
%----------------------------------------------------------------------------------------------------------------
\begin{eqnarray}
{\cal D}_x^{q}
G\left(\left\{f_b(\tau),f_{a_1}(\tau),...f_{a_k}(\tau)\right\};x\right)_b
= \left\{\frac{d}{dx}
G\left(\left\{f_b(\tau),f_{a_1}(\tau),...f_{a_k}(\tau)\right\};x\right)\right\}_q. 
\end{eqnarray}
%----------------------------------------------------------------------------------------------------------------
The definition of the $q$-free iterated integrals being considered in the following has 
been given in Ref.~\cite{Blumlein:2026clg}.
The iterated integrals obey $q$-differential equations, which can be derived 
algorithmically.
The order of these equations rises with the weight of the corresponding iterated 
integrals.

Now we turn to the individual classes of iterated integrals: the 
polylogarithm \cite{LEWIN1,LEWIN2,Devoto:1983tc}, 
the Nielsen integrals \cite{NIELSEN1,Kolbig:1983qt},
harmonic polylogarithms \cite{Remiddi:1999ew}, 
generalized harmonic polylogarithms, also called Kummer-Poincar\'e integrals 
\cite{KUMMER1,KUMMER2,KUMMER3,POINCARE1,LAPPO,CHEN,GONCHAROV,Moch:2001zr,Ablinger:2013cf}, 
cyclotomic iterated integrals \cite{Ablinger:2011te},
iterated integrals implied by quadratic forms \cite{Ablinger:2021fnc}, 
and iterated integrals containing square-root valued letters \cite{Ablinger:2014bra}.

%----------------------------------------------------------------------------------------------------------------
\subsection{The \boldmath $q$-extended polylogarithm}  
\label{sec:41}
%----------------------------------------------------------------------------------------------------------------

\vspace*{1mm}
\noindent
The polylogarithms 
\cite{LEWIN1,LEWIN2,Devoto:1983tc}
are iterated integrals over the alphabet 
%----------------------------------------------------------------------------------------------------- 
\begin{eqnarray} 
\label{eq:ALPHA:P} 
\mathfrak{A}_P = 
\left\{f_0(x), f_1(x)\right\} =  
\left\{\frac{1}{x},\frac{1}{1-x}\right\}, 
\end{eqnarray} 
%----------------------------------------------------------------------------------------------------- 
as in terms of harmonic polylogarithms \cite{Remiddi:1999ew} (see also 
Section~\ref{sec:43}),
%----------------------------------------------------------------------------------------------------- 
\begin{eqnarray} 
\Li_k(x) = \HA_{\tiny {\underbrace{0,...,0}_{k-1},1}}(x), 
\end{eqnarray} 
%----------------------------------------------------------------------------------------------------- 
with
%----------------------------------------------------------------------------------------------------- 
\begin{eqnarray} 
\Li_k(x) = \sum_{n = 1}^\infty \frac{x^n}{n^k},~~k \geq 1,~~x \in [-1,1]. 
\end{eqnarray} 
%----------------------------------------------------------------------------------------------------- 
The function $f[n]$ for polylogarithms $\Li_k(x)$  is
%----------------------------------------------------------------------------------------------------------------
\begin{eqnarray} 
f[n] = \frac{1}{n^k},~~~k \in \mathbb{Z},~~~n \in \mathbb{N} \backslash \{0\}. 
\end{eqnarray}
%----------------------------------------------------------------------------------------------------------------
The polylogarithms for non-positive index $k$ are given by
%----------------------------------------------------------------------------------------------------------------
\begin{eqnarray} 
\Li_{0}(x) &=& \frac{x}{1-x},
\\
\Li_{-k}(x) &=& \frac{1}{(1-x)^{k+1}} \sum_{l=0}^{k-1} A_{k,l} x^{k-l},~~k \geq 1,
\end{eqnarray}
%----------------------------------------------------------------------------------------------------------------
where $A_{k,l}$ are the Eulerian numbers \cite{EULER5}
%----------------------------------------------------------------------------------------------------------------
\begin{eqnarray} 
A_{k,l} = (k-l) A_{k-1,l-1} + (k+1) A_{n-1,k},~~A_{0,0} = 1,~A_{0,k} = 0, k \neq 0.
\end{eqnarray}
%----------------------------------------------------------------------------------------------------------------

Polylogarithms for $k > 0$ obey homogeneous differential equations of order $k+1$. The 
first in the series are
%----------------------------------------------------------------------------------------------------------------
\begin{eqnarray} 
\left[-(1-x) \frac{d^2}{dx^2} + \frac{d}{dx}\right] \Li_1(x) &=& 0, 
\\
\left[-(1-x) x \frac{d^3}{dx^3} + (-2 +3 x) \frac{d^2}{dx^2} + \frac{d}{dx}
\right] \Li_2(x) &=& 0, 
\\
\left[-(1-x) x^2 \frac{d^4}{dx^4} + (-5 + 6 x) x \frac{d^3}{dx^3} + (-4 + 
7 x) \frac{d^2}{dx^2} + \frac{d}{dx}
\right] \Li_3(x) &=& 0. 
\end{eqnarray}
%----------------------------------------------------------------------------------------------------------------

The $q$-extended polylogarithms are defined by
%----------------------------------------------------------------------------------------------------------------
\begin{eqnarray} 
\Li_k(x;q) = \sum_{n=1}^\infty \frac{x^n}{\{n\}_q^k} 
= \sum_{n=1}^\infty x^n \left[\frac{(1-q)}{(1-q^n)}\right]^k.  
\end{eqnarray}
%----------------------------------------------------------------------------------------------------------------
This representation also allows one to define the $q$-$\zeta$ values, the extensions of the 
Riemann-$\zeta$ 
values,\footnote{In other approaches additional $q$-numerator powers are considered.
This allows one to expand in $q$ because of the condition Eq.~(\ref{eq:BAS1}), 
cf.~Ref.~\cite{KUEHN}. In our representation for infinite sums the 
limit $q \rightarrow 0$ has to be excluded.}  
%----------------------------------------------------------------------------------------------------------------
\begin{eqnarray} 
\lim_{x \rightarrow 1^-} \Li_k(x;q) = \zeta_k(q) = \sum_{n=1}^\infty 
\frac{1}{\{n\}_q^k} 
= \sum_{n=1}^\infty \left[\frac{(1-q)}{(1-q^n)}\right]^k,~~~k 
\geq 2,~~q \in (0,1],  
\end{eqnarray}
%----------------------------------------------------------------------------------------------------------------
unlike the case for the $\mu$-extension in 
Ref.~\cite{Blumlein:2026clg}.\footnote{Note, however, that a different 
version of $\mu$-extension has been used in Refs.~\cite{Rebesh:2013hps,50a}.
There the values of $\mu$-$\zeta$ exist and are used in determining the critical 
temperature of $\mu$ Bose-gas condensation.} 
The usual $\zeta$-values are 
obtained in the limit $q \rightarrow 1^-$.

By applying the $q$-derivative to $\Li_k(x;q)$ one obtains
%----------------------------------------------------------------------------------------------------------------
\begin{eqnarray} 
x {\cal D}_x^ q \Li_k(x;q) = \Li_{k-1}(x;q),  
\end{eqnarray}
%----------------------------------------------------------------------------------------------------------------
with
%----------------------------------------------------------------------------------------------------------------
\begin{eqnarray} 
\Li_{1}(x;q) = - \ln(1-x)|_q = \sum_{n=1}^\infty x^n \frac{(1-q)}{(1-q^n)} 
\end{eqnarray}
%----------------------------------------------------------------------------------------------------------------
and for non-positive values of $k$ one can resum the infinite series and obtains
%----------------------------------------------------------------------------------------------------------------
\begin{eqnarray} 
\Li_{0}(x;q) &=& \frac{x}{(1-x)} \equiv \Li_0(x), \\
\Li_{-1}(x;q) &=& \frac{x}{(1-x)(1-q x)},  \\
\Li_{-2}(x;q) &=& \frac{x(1 + qx)}{(1-x)(1-q x)(1-q^2 x)},  \\
\Li_{-3}(x;q) &=& \frac{x (1 + 2 q x + 2 q^2 x + q^3 x^2)}{(1 - x) (1 - q x) (1 - 
   q^2 x) (1 - q^3 x)}, 
\end{eqnarray}
%----------------------------------------------------------------------------------------------------------------
etc.

For $k > 0$ one obtains the following representation in terms of $q$-extended
generalized hypergeometric functions 
%----------------------------------------------------------------------------------------------------------------
\begin{eqnarray} 
\label{eq:Likhyp}
\Li_k(x;q) &=&
x \cdot {_{k+1}\phi_{k}} \left( 
  \begin{matrix} 
    q, \dots, q \\ 
    q^2,  \dots, q^2 
  \end{matrix} 
  ; q, x 
\right) 
 = 
\sum_{n=0}^{\infty}
\frac{\small {\overbrace{(q; q)_n \cdots (q; q)_n}^{k}}}
     {\small {\underbrace{(q^2; q)_n \cdots (q^2; q)_n}_{k}}} x^{n+1}; k \geq 1. 
\end{eqnarray}
%----------------------------------------------------------------------------------------------------------------
The recurrence of the expansion coefficients of the $q$-polylogarithms are
%----------------------------------------------------------------------------------------------------------------
\begin{eqnarray}
(1 - q^{n})^k f[n;q] - (1 - q^{n+1)})^k f[n+1;q] = 0;~~f[1;q] = 1,~~k \geq 1. 
\end{eqnarray}
%----------------------------------------------------------------------------------------------------------------

Examples of the differential equations for $\Li_k(x;q)$ are\footnote{Based on the $q$-recurrences one may find the associated
$q$-differential equations by the command {\tt qREToqDE} and the $q$-shift relations by {\tt qREToqSE}, 
Ref.~\cite{UNCU} or by related commands in Ref.~\cite{KOUTSCHAN}.} 
%----------------------------------------------------------------------------------------------------------------
\begin{eqnarray} 
\left((q x-1) {\cal D}_x^{q} + 1\right) \Li_1(x;q) &=& \frac{1}{x-1},\nonumber\\
\Li_1(0;q) = 0, &&
\\
\left(q x \left(q^2 x-1\right) ({\cal D}_x^{q})^2 +(q (q+2)
   x-1) {\cal D}_x^{q} + 1\right) \Li_2(x;q) &=& \frac{1}{x-1}, \nonumber\\
 \Li_2(0;q) = 0,~~~{\cal D}_x^{q} \Li_2(0;q) = 1 &&
\\
\biggl[q x \left(q^2 x \left(q^3 x \left(q^3 x-1\right)
   ({\cal D}_x^{q})^3+(q ((q (q (q+2)+3)+3) q x-q-2)-3)
   ({\cal D}_x^{q})^2\right)  \right. 
&& \nonumber \\ \left. 
+(q (q (q (q (q+3)+6)+6)+3) x-q (q+3)-3)
   {\cal D}_x^{q}\right) + \left((q+1)^3 x-1\right) \biggr]\Li_3(x;q) &=& 
\frac{1}{x-1},
\nonumber\\ 
\Li_3(0;q)=1,~{\cal D}_x^{q} \Li_3(0;q)=1 + \frac{1}{(q+1)^3}, 
&& \nonumber\\
~({\cal D}_x^{q})^2 \Li_3(0;q) = 
1
+q
+\frac{1}{(q+1)^2} +
   \frac{q+1}{\left(q^2+q+1\right)^3}. &&
\end{eqnarray}
%----------------------------------------------------------------------------------------------------------------
The differential equations also apply to Eq.~(\ref{eq:Likhyp}).

The corresponding $q$-shift relations for $\Li_k(x;q),~~k=1,2,3,$ are given by
%----------------------------------------------------------------------------------------------------------------
\begin{eqnarray} 
(1-x) (q x-1) F[q x] + (x-1)^2 F[x]+(q-1) x &=& 0,
\\
(1-x) \left(q^2 x-1\right) F[q^2 x] + 2 (x-1) (q x-1)
   F[q x]-(x-1)^2 F[x]+(q-1)^2 x &=& 0,
\\
q^3 (x-1) \left(q^3 x-1\right) F\left(q^3 x\right)-3 q^2 (x-1)
   \left(q^2 x-1\right) F\left(q^2 x\right)
\nonumber\\ 
+3 q (x-1) (q x-1)
   F(q x)-(x-1)^2 F(x)-(q-1)^3 &=& 0.
\end{eqnarray}
%----------------------------------------------------------------------------------------------------------------
Quantum dilogarithms were considered in the 1990s in Refs.~\cite{FS93,Faddeev:1993rs}.
The $q$-logarithm dates back to Euler \cite{EULER2}. Compared to our definition of $\Li_k(x;q)$,
deviating representations are used in the literature.
Ref.~\cite{Kirillov:1994en} uses a variant of the $q$-dilogarithm defined by
%-----------------------------------------------------------------------------------------------------
\begin{eqnarray} 
\tilde{\Li}_2(x;q) = \sum_{n=1}^\infty \frac{x}{n} \frac{1-q}{1-q^n}, 
\end{eqnarray} 
%-----------------------------------------------------------------------------------------------------
and analogously for the polylogarithm in Ref.~\cite{KIRILLOV2}. Our convention has been used in 
Ref.~\cite{SCHLESINGER} for the multiple polylogarithm \cite{Borwein:1999js}, 
up to factors of $(1-q)^k$.
In Ref.~\cite{Goncharov:2026zqy} the $q$-multiple polylogarithm \cite{GON01} is defined using the 
symmetric $q$ extension, see Ref.~\cite{Blumlein:2026clg}, Appendix~A, up to factors 
$(1-q)^k$.
Further powers of $q$ are introduced in the numerators of the summands
in Ref.~\cite{BRADLEY} for the related $q$-multiple zeta values, see also Refs.~\cite{KUEHN,ZHAO,OKUDA,REINEKE}. 
%-------------------------------------------------------------------------------------

%-----------------------------------------------------------------------------------------------------
\subsection{The \boldmath $q$-deformed Nielsen integrals}
\label{sec:42}
%-----------------------------------------------------------------------------------------------------

\vspace*{1mm}
\noindent 
The Nielsen integrals \cite{NIELSEN1,Kolbig:1983qt}
are the iterative integrals
%-----------------------------------------------------------------------------------------------------
\begin{eqnarray} 
S_{n,p}(x) = \HA_{\tiny {\underbrace{0,...,0}_n,\underbrace{ 
1,...,1}_p}}(x), 
\end{eqnarray} 
%-----------------------------------------------------------------------------------------------------
over $\mathfrak{A}_P$ written as a harmonic polylogarithm. They can be represented by
%-----------------------------------------------------------------------------------------------------
\begin{eqnarray}
\label{eq:SPN3}
S_{n,p}(x) &=& \sum_{k=p}^\infty \left[\begin{array}{c} k \\ p \end{array} \right]    
\frac{x^k}{k^{n+1}} 
\end{eqnarray} 
%-----------------------------------------------------------------------------------------------------
by their expansion around $x = 0$.
Here the Stirling numbers of the first kind in the 
notation of Refs.~\cite{COMTET,Kolbig:1983qt,ADAMCHIK1} are expressed
as modified Faa di Bruno-Ramanujan 
determinants \cite{FDB,RAMAN,Blumlein:1998if}
%-----------------------------------------------------------------------------------------------------
\begin{eqnarray}
\label{eq:stir}
\left[\begin{array}{c} k \\ p \end{array} \right]
&=& \frac{1}{(p-1)!} \left| \begin{array}{cccccccc}
S_1(k-1) & 1     & 0 &      0 & 0  & ... & 0\\
S_2(k-1) & S_1(k-1)& 2 &      0 & 0  & ... & 0\\
S_3(k-1) & S_2(k-1)& S_1(k-1) & 3 & 0  & ... & 0\\
\vdots &       &        &   &    &     & \vdots  \\
S_{p-1}(k-1) & ...      &        &   &    & ... &S_1(k-1)  \\
\end{array} \right|,~~p > 1;
\nonumber\\
\\
\left[\begin{array}{c} k \\ 1 \end{array} \right] &=& (k-1)!,~~~ 
\left[\begin{array}{c} k \\ 0 \end{array} \right] = \delta_{0k}, 
\end{eqnarray}
%-----------------------------------------------------------------------------------------------------
with 
%-----------------------------------------------------------------------------------------------------
\begin{eqnarray}
\left[\begin{array}{c} k \\ p \end{array} \right] = s(k,p) \frac{(-1)^k}{(k-1)!}.
\end{eqnarray}
%-----------------------------------------------------------------------------------------------------

Furthermore, one has
%-----------------------------------------------------------------------------------------------------
\begin{eqnarray}
S_{0,p} = \frac{1}{p!} \ln^p(1-x). 
\end{eqnarray}
%-----------------------------------------------------------------------------------------------------
Nielsen integrals of weight ${\sf w} = n + p$ obey differential equations of order 
$n + p + 1$. Examples are
%-----------------------------------------------------------------------------------------------------
\begin{eqnarray}
&& \Biggl[
  x (x-1)^2 \frac{d^4}{dx^4} 
+ 3 (2 x-1)(x-1) \frac{d^3}{dx^3}
+ (7 x-6) \frac{d^2}{dx^2}
+ \frac{d}{dx}
\Biggr] S_{1,2}(x) = 0,  \\
%----
&& \Biggl[
 x (x-1)^3 
\frac{d^5}{dx^5} 
+ 2 (5 x-2) (x-1)^2 \frac{d^4}{dx^4} 
+ (25 x-18)
   (x-1)
   \frac{d^3}{dx^3} 
+  (15
   x-14)
   \frac{d^2}{dx^2} 
+ \frac{d}{dx}
\Biggr] \nonumber\\ && \times
 S_{1,3}(x) = 0, 
\\
&& \Biggl[
   (1-x)^2 x^2 \frac{d^5}{dx^5}
+  (x-1) (10 x-7) x \frac{d^4}{dx^4}
+ \left(25 x^2-33 x+9\right) \frac{d^3}{dx^3}
+3 (5 x-4) \frac{d^2}{dx^2}
+ \frac{d}{dx}
\Biggr] 
\nonumber\\ && \times
S_{2,2}(x) = 0. 
\end{eqnarray}
%-----------------------------------------------------------------------------------------------------

The $q$-extension $S_{n,p}(x;q)$ is obtained by
%-----------------------------------------------------------------------------------------------------
\begin{eqnarray}
\label{eqSNP}
S_{n,p}(x;q) &=& \sum_{k=p}^\infty \left[\begin{array}{c} k \\ p \end{array} \right]_q
\left[\frac{(1-q)}{(1-q^k)}\right]^{n+1} x^k,
\end{eqnarray}
%-----------------------------------------------------------------------------------------------------
with
%-----------------------------------------------------------------------------------------------------
\begin{eqnarray}
\left[\begin{array}{c} k \\ p 
\end{array} \right]_q =
\left[\begin{array}{c} k \\ p \end{array} \right]\{S_l(k-1) \rightarrow 
S_l(k-1;q)\},~~l \geq 1, l \in \mathbb{N}.
\end{eqnarray}
%-----------------------------------------------------------------------------------------------------
The hierarchy equations for the $q$-extended Nielsen integrals  are 
%-----------------------------------------------------------------------------------------------------
\begin{eqnarray}
x {\cal D}_x^{(q)} S_{n,p}(x;q) &=& 
S_{n-1,p}(x;q). 
\end{eqnarray}
%-----------------------------------------------------------------------------------------------------
The expansion coefficients $f[n;q]$ for $S_{1,2}(x;q), S_{1,3}(x;q)$ and  $
S_{2,2}(x;q)$ obey the recurrences
%-----------------------------------------------------------------------------------------------------
\begin{eqnarray}
%% S_{1,2}
&& -\big(
        1-q^n\big)^3 f[n;q]
-\big(
        1-q^{1
        +n
        }\big)^2 \big(
        -2
        +q^n
        +q^{1+n}
\big) f[n+1,q]
-\big(
        1-q^{1
        +n
        }
\big)
\big(1-q^{2
        +n
        }\big)^2 
\nonumber\\ && \times f[n+2;q] = 0, 
\\ 
%% S_{1,3}
&& 
\left(q^n-1\right)^4
   f[n;q]
-\left(q^{n+1}-1\right)^2 \left(-3 (q+1)
   q^n+\left(q^2+q+1\right)
   q^{2 n}+3\right)
   f[n+1;q]
\nonumber\\ &&
+\left(q^{n+1}-1\right) \left(q^{n+2}-1\right)^2
   \left(\left(q^2+q+1\right)
   q^n-3\right) f[n+2;q]
+  
\left(1-q^{n+1}\right)
\left(q^{n+2}-1\right)
\nonumber\\ && \times
   \left(q^{n+3}-1\right)^2
   f[n+3;q] = 0, 
\\
%% S_{2,2}
&& \big(
        1-q^n\big)^4 f[n;q] 
+\big(
        1-q^{1
        +n
        }
\big)
\biggl[\big(
                1-q^{1
                +n
                }\big)^2 
\big(
                -2
                +q^n
                +q^{1+n}
        \big) f[n+1;q]
\nonumber\\ &&
        +\big(
                1-q^{2
                +n
                }\big)^3 f[n+2;q]
\biggr]
= 0. 
%---
\end{eqnarray}
%-----------------------------------------------------------------------------------------------------
The initial conditions are obtained by the first expansion 
coefficients of (\ref{eqSNP}).
%-----------------------------------------------------------------------------------------------------
The $q$-differential equations for 
$S_{1,2}(x;q),
S_{1,3}(x;q)$ and 
$S_{2,2}(x;q)$ are 
%-----------------------------------------------------------------------------------------------------
\begin{eqnarray}
&& 
\biggl[
q^3 x
  \left(q^2 x-1\right)
({\cal D}_x^{(q)})^4
+(q x-1) \left(q^4
   x+2 q^3 x+q^2 (3
   x-1)-q-1\right)
({\cal D}_x^{(q)})^3
\nonumber\\ &&
+\left(q^3 x+q^2 (3 x-1)+3 q
   (x-1)-2\right) 
({\cal D}_x^{(q)})^2
+ {\cal D}_x^{(q)}
\biggr] F[x] = 0
\nonumber\\ &&
F[0] = 
{\cal D}_x^{(q)}F[0] = 0,
({\cal D}_x^{(q)})^2 F[0] = \frac{1}{1+q},
({\cal D}_x^{(q)})^3 F[0] = \frac{2 + q}{1 + q + q^2}.
\\
%----------
&& 
\biggl[
q^4 x
   \left(q^3 x-1\right) (q^2 x -1) 
({\cal D}_x^{(q)})^5 
+\left(q^2
   x-1\right) \left(q^6
   x+2 q^5 x+3 q^4 x+q^3 (4
   x-1)-q^2 \right.
\nonumber\\ && \left. 
-q-1\right)
({\cal D}_x^{(q)})^4 
+(q x-1) \left(q^6 x+3
   q^5 x+q^4 (7 x-1)+q^3 (8
   x-3)+6 q^2 (x-1)-5
   q-3\right)
\nonumber\\ && 
\times ({\cal D}_x^{(q)})^3
+\left(q^4 x+q^3 (4
   x-1)+q^2 (6 x-4)+q (4
   x-6)-3\right)
({\cal D}_x^{(q)})^2
+ 
{\cal D}_x^{(q)})\biggr] F[x] = 0,
\nonumber\\
&& F[0]~=~ 
{\cal D}_x^{(q)} F[0]~=~ 
({\cal D}_x^{(q)})^2 F[0]~=~0,
({\cal D}_x^{(q)})^3 F[0]~=~\frac{2}{1 + q + q^2},
\nonumber\\ &&
({\cal D}_x^{(q)})^4 F[0]~=~\frac{2 (3 + 2 q + q^2)}{(1 + q) (1 + q^2)},
\\
%%---
&&
\biggl[
q^7 x^2
   (q x-1) \left(q^2
   x-1\right)
({\cal D}_x^{(q)})^5
+q^3 x (q
   x-1) \left(q^5 x+2 q^4
   x+q^3 (3 x-1)+q^2 (4 x-2) \right. 
\nonumber\\ &&
\left.
-2
   q-2\right)
({\cal D}_x^{(q)})^4
+\left(q^7 x^2+q^6 x (3
   x-1)+q^5 x (7 x-4)+q^4
   \left(8 x^2-9
   x+1\right) \right.
\nonumber\\ && \left.
+q^3 \left(6
   x^2-11 x+2\right)+q^2 (3-7
   x)-q (x-2)+1\right)
({\cal D}_x^{(q)})^3
+(q+2)
   \left(q^3 x+q^2 (2 x-1) \right.
\nonumber\\ && \left.
+2 q
   (x-1)-1\right)
({\cal D}_x^{(q)})^2
+ {\cal D}_x^{(q)}
\biggr] 
F(x) = 0, 
\nonumber\\ &&
F(0) = 
{\cal D}_x^{(q)} F(0) = 0,
({\cal D}_x^{(q)})^2 F(0) = \frac{1}{(1+q)^2},
({\cal D}_x^{(q)})^3 F(0) = \frac{2 + q}{(1 + q + q^2)^2},
\nonumber\\ &&
({\cal D}_x^{(q)})^4 F(0) = \frac{3 + 4 q + 3 q^2 + q^3}{(1 + q)^2 (1 + q^2)^2}.
\end{eqnarray}
%-----------------------------------------------------------------------------------------------------
The associated $q$-shift relations for $S_{1,2}(x;q), S_{1,3}(x;q)$ and  $S_{2,2}(x;q)$ are
%-----------------------------------------------------------------------------------------------------
\begin{eqnarray} 
&& 
q^3 (x-1)^2 F[x]
-q \left(q^2
   \left(4 x^2-5 x+2\right)-3
   q x+q+1\right) F[q x]
+\left(q^3 \left(6
   x^2-4 x+1\right) \right.
\nonumber\\ && \left.
+q^2 (2-7
   x)-q (x-2)+1\right)
   F[q^2 x]
-(q x-1)
   \left(q^2 (4
   x-1)-q-2\right) F[q^3 x]
\nonumber\\ &&
+(q
   x-1) \left(q^2 x-1\right)
   F[q^4 x] = 0,  
\\
&& 
-q^6 (x-1)^3
   F(x)
+q^3 \left(q^3 \left(5
   x^3-9 x^2+7 x-2\right)+q^2
   \left(-6 x^2+4 x-1\right)+q
   (4 x-1)-1\right) 
\nonumber\\ &&
\times
F[q x]
-q \left(q^5
   \left(10 x^3-10 x^2+5
   x-1\right)+q^4 \left(-16
   x^2+9 x-2\right)-2 q^3     \left(2 x^2-5
   x+1\right)
\right.
\nonumber\\ &&  \left.
+q^2 (5 x-3)+q
   (x-1)-1\right) 
   F[q^2 x] 
+(q x-1)
   \left(q^5 \left(10 x^2-5
   x+1\right)+q^4 (1-5 x)+q^3
   (3-9 x) \right.
\nonumber\\ && \left.
-q^2 (x-2)+2
   q+1\right) 
   F[q^3 x]
-(q x-1)
   \left(q^2 x-1\right)
   \left(q^3 (5
   x-1)-q^2-q-2\right)
   F[q^4 x]
\nonumber\\ &&
+(q
   x-1) \left(q^2 x-1\right)
   \left(q^3 x-1\right)
   F[q^5 x]
= 0, 
\\
&&
-q^3 (x-1)^2
   F[x]
+q \left(q^2 \left(5
   x^2-7 x+3\right)-3 q
   x+q+1\right) F[q x]
+  \left(q^3
   \left(-10 x^2+9
   x-3\right) \right.
\nonumber\\ && \left.
+q^2 (10 x-3)+q
   (x-3)-1\right) 
F[q^2 x]
+  \left(q^3 \left(10
   x^2-5 x+1\right)+q^2 (3-12
   x)-3 q (x-1)+3\right)
\nonumber\\ && \times
   F[q^3 x]
-(q x-1)
   \left(q^2 (5
   x-1)-q-3\right) 
F[q^4 x]
+(q x-1)
   \left(q^2 x-1\right)
   F[q^5 x]
= 0. 
\end{eqnarray}
%-----------------------------------------------------------------------------------------------------
%-----------------------------------------------------------------------------------------------------
\subsection{The \boldmath $q$-deformed harmonic polylogarithms}
\label{sec:43}
%-----------------------------------------------------------------------------------------------------

\vspace*{1mm}
\noindent 
The harmonic polylogarithms \cite{Remiddi:1999ew} are iterated integrals over the 
alphabet
%-----------------------------------------------------------------------------------------------------
\begin{eqnarray}
\mathfrak{A}_{\rm H} = 
\left\{f_0(x), f_1(x), f_{-1}(x)\right\} \equiv 
\left\{\frac{1}{x}, \frac{1}{1-x}, \frac{1}{1+x}\right\}, 
\end{eqnarray} 
%-----------------------------------------------------------------------------------------------------
by
%-----------------------------------------------------------------------------------------------------
\begin{eqnarray}
\HA_{b,\vec{a}}(x) = \int_0^x dy f_b(y) \HA_{\vec{a}}(y),~~~\HA_\emptyset(x) = 1.  
\end{eqnarray}
%-----------------------------------------------------------------------------------------------------
The most simple harmonic polylogarithms are $\HA_{-1}(x), \HA_{1}(x), \HA_{0}(x)$.
The first two have the following $f[n]$-function
%-----------------------------------------------------------------------------------------------------
\begin{eqnarray} 
f[n]^{(-1)}  &=& \frac{(-1)^n}{n},  \\
f[n]^{(1)}   &=& \frac{1}{n}, 
\end{eqnarray} 
%-----------------------------------------------------------------------------------------------------
from which
%-----------------------------------------------------------------------------------------------------
\begin{eqnarray} 
\HA_{-1}(x;q)  &=&  \sum_{n=1}^\infty \frac{1 - q}{1 - q^k} (-x)^k
= 
x \cdot {_2\phi_1} \left( 
  \begin{matrix} 
    q, q \\ 
    q^2
  \end{matrix} 
  ; q, -x
\right),  \\
\HA_{ 1}(x;q)  &=&  \sum_{n=1}^\infty \frac{1 - q}{1 - q^k} x^k
= 
x \cdot {_2\phi_1} \left( 
  \begin{matrix} 
    q, q \\ 
    q^2
  \end{matrix} 
  ; q, x
\right) 
\end{eqnarray} 
%-----------------------------------------------------------------------------------------------------
are obtained. This yields the representation
%-----------------------------------------------------------------------------------------------------
\begin{eqnarray} 
\HA_{0}(x;q)  &=&  -\sum_{n=1}^\infty \frac{1 - q}{1 - q^k} (1-x)^k. 
\end{eqnarray} 
%-----------------------------------------------------------------------------------------------------

Due to the complexity of the iterative integrals by growing alphabets and permutation 
of letters in this and the following cases, we will refer to characteristic examples 
henceforth.
Let us consider the harmonic polylogarithm $\HA_{-1,0,1}(x)$. Its function $f[n]$
is given by
%-----------------------------------------------------------------------------------------------------
\begin{eqnarray}
\label{eq:fnH}
f[n] =\left[\frac{1}{n^3} - \frac{(-1)^n}{n} S_{-2}(n)\right]. 
\end{eqnarray}
%-----------------------------------------------------------------------------------------------------
The differential equation for  $\HA_{-1,0,1}(x)$ reads
%-----------------------------------------------------------------------------------------------------
\begin{eqnarray}
\left[-(1 - x) x (1 + x) \frac{d^4}{dx^4} + 2 (-1 - x + 3 x^2) \frac{d^3}{dx^3}
+ (-3 + 7 x) \frac{d^2}{dx^2}  + \frac{d}{dx}\right] \HA_{-1,0,1}(x) = 0. 
\end{eqnarray}
%-----------------------------------------------------------------------------------------------------

The $q$-extension of Eq.~(\ref{eq:fnH}) is given by
%-----------------------------------------------------------------------------------------------------
\begin{eqnarray}
f[n;q] =(1-q)^3 \Biggl[
\frac{1}{(1-q^n)^3} - \frac{(-1)^n}{(1-q^n)} 
\sum_{k=1}^n \frac{(-1)^k}{(1-q^k)^2}\Biggr]. 
\end{eqnarray}
%-----------------------------------------------------------------------------------------------------
It obeys the recurrence
%-----------------------------------------------------------------------------------------------------
\begin{eqnarray}
&& 
-\left(q^n-1\right)^3 f[n,q]
+(q-1)
   \left(q^{n+1}-1\right)
   \left(q^{n+1}+q^n-2\right) q^n f[n+1;q]
\nonumber\\ &&
+\left(q^{n+1}-1\right)^2 \left(q^{n+2}-1\right)
   f[n+2;q] = 0.  
\end{eqnarray}
%-----------------------------------------------------------------------------------------------------

The $q$-extension of the derivative of $\HA_{-1,0,1}(x)$,
$\HA_{0,1}(x)/(1+x)$, is
%---------------------------------------------------------------------------------------------$
\begin{eqnarray}
f^{(1)}[n;q] =(-1)^n \sum_{k=1}^n (-1)^k \frac{(1-q)^2}{(1-q^k)^2}. 
\end{eqnarray} 
%-----------------------------------------------------------------------------------------------------
Here the hierarchy-equation is
%-----------------------------------------------------------------------------------------------------
\begin{eqnarray}
{\cal D}_x^{(q)} \HA_{-1,0,1}(x;q) = \left[\frac{1}{1+x} 
\left. \HA_{0,1}(x)\right]\right|_q. 
\end{eqnarray}
%-----------------------------------------------------------------------------------------------------

The $q$-extended harmonic polylogarithm $\HA_{-1,0,1}(x)$ obeys
%-----------------------------------------------------------------------------------------------------
\begin{eqnarray}
&& \biggl[{\cal D}_x^{(q)}
+q
   \left[q^2 x+q (3 x-1)+3
   x-2\right]
   ({\cal D}_x^{(q)})^2
+\left[q^5 x^2+q^4 x (2
   x-1)+q^3 x (3 x-2) \right.
\nonumber\\ && \left.
-q^2 x
+q
   (2 x-1)-1\right]
   ({\cal D}_x^{(q)})^3
+q^2 x (q
   x-1) \left(q^3 x+1\right)
   ({\cal D}_x^{(q)})^4 \biggr] F[x]
= 0, 
\end{eqnarray}
%-----------------------------------------------------------------------------------------------------
with 
%-----------------------------------------------------------------------------------------------------
\begin{eqnarray}
F[0]~=~{\cal D}_x^{(q)}~F[0]~= 0,~~
({\cal D}_x^{(q)})^2~F[0] = 1,~~
({\cal D}_x^{(q)})^3~F[0] =-\frac{q (q+2)}{q+1}.  
\end{eqnarray}
%-----------------------------------------------------------------------------------------------------
%-----------------------------------------------------------------------------------------------------
The $q$-shift relation is given by
%-----------------------------------------------------------------------------------------------------
\begin{eqnarray}
&& q^4 (x-1) (x+1) F[x]
  +q^2 \left(q^2 \left(-4 x^2+2 x+1\right)-2 q (x-1)+1\right) F[q x]
\nonumber\\ &&
  +q \left(3 q^3 x (2 x-1)-2 q^2+q (3 x-2)-2\right) F\left[q^2 x\right]
  +\left(q^4 \left(x-4 x^2\right)+3 q^3 x+q^2 (1-3 x) \right.
\nonumber\\ && \left.
-q
   (x-2)+1\right) F\left[q^3 x\right] 
  +(q x-1) \left(q^3 x+1\right) F\left[q^4 x\right]
= 0. 
\end{eqnarray}
%-----------------------------------------------------------------------------------------------------
Similar relations hold for all $q$-extended harmonic polylogarithms.

The $q$-multiple zeta values can be obtained from the $q$-extended harmonic 
polylogarithms for $x=1$
like for the multiple zeta values \cite{Blumlein:2009cf} from the harmonic 
polylogarithms,
%-----------------------------------------------------------------------------------------------------
\begin{eqnarray}
\zeta_{-1,0,1}(q) = \HA_{-1,0,1}(1;q) = \sum_{k=1}^\infty f[k;q]. 
\end{eqnarray}
%-----------------------------------------------------------------------------------------------------
In the limit $q \rightarrow 1^-$, $\zeta_{-1,0,1}(q)$ 
approaches 
%-----------------------------------------------------------------------------------------------------
\begin{eqnarray}
\HA_{-1,0,1}(1) = \ln(2) \zeta_2 - \frac{5}{8} \zeta_3.  
\end{eqnarray}
%-----------------------------------------------------------------------------------------------------
%-----------------------------------------------------------------------------------------------------
\subsection{The \boldmath $q$-deformed generalized harmonic polylogarithms}
\label{sec:44}
%-----------------------------------------------------------------------------------------------------

\vspace*{1mm}
\noindent
We consider the $q$-extension of the generalized harmonic polylogarithms, cf.~Refs.~
\cite{Moch:2001zr,Ablinger:2013cf}. The function
$\HA[-1/2,2,-1,x]$ is a typical example. We calculate it algorithmically. The 
associated
function $f[n]$ reads
%-----------------------------------------------------------------------------------------------------
\begin{eqnarray}
\label{eq:GHS1}
f[n] = \frac{(-1)^n}{n^3}
-\frac{2^{-n}}{n^2} S_1({{-2},n})
-\frac{(-2)^n}{n} S_2\left({{\frac{1}{2}},n}\right)
+\frac{(-2)^n}{n} S_{1,1}\left({{-\frac{1}{4},-2},n}\right), 
\end{eqnarray}
%-----------------------------------------------------------------------------------------------------
cf.~Ref.~\cite{Blumlein:2026clg}. The $q$-extension is
%-----------------------------------------------------------------------------------------------------
\begin{eqnarray}
\label{eq:GHS2}
\lefteqn{f[n;q] = } \nonumber \\ && 
(-1)^n \left(\frac{1-q}{1-q^n}\right)^3
-2^{-n} \left(\frac{1-q}{1-q^n} \right)^2 \sum_{k=1}^n (-1)^k 
\left(\frac{1-q}{1-q^k}\right)
\nonumber\\ &&
-(-2)^n \left(\frac{1-q}{1-q^n}\right)
\sum_{k=1}^n \left(\frac{1}{2}\right)^k 
\left(\frac{1-q}{1-q^k}\right)^2
+(-2)^n \left(\frac{1-q}{1-q^n}\right) 
\nonumber\\ && \times
\sum_{k=1}^n \left(-\frac{1}{4}\right)^k 
\left(\frac{1-q}{1-q^k}\right) \sum_{l=1}^k (-2)^l \left(\frac{1-q}{1-q^l}\right)
\nonumber\\ &=&
(-2)^n \left(\frac{1-q}{1-q^n}\right) 
\sum_{k=1}^{n-1} 
\Biggl\{
\left(-\frac{1}{4}\right)^k 
\left(\frac{1-q}{1-q^k}\right) \sum_{l=1}^k (-2)^l \left(\frac{1-q}{1-q^l}\right)
-
\left(\frac{1}{2}\right)^k 
\left(\frac{1-q}{1-q^k}\right)^2
\Biggr\}.	 \nonumber\\ 
\end{eqnarray}
%-----------------------------------------------------------------------------------------------------
The function $F[x;q]$, cf.~Eq.~(\ref{eq:2}), starts $\propto x^3$.

Here and in the following it is useful to determine the recurrences for the individual 
terms first. Afterwards these recurrences are combined into a single 
recurrence. Furthermore, the use of formal parameters for the power-terms is usually 
faster then considering the special ones. Therefore one sets, e.g., $(-2)^k \equiv a^k$, 
etc. This also avoids complex algebra in some of the examples discussed below. 

Let us consider the example 
%-----------------------------------------------------------------------------------------------------
\begin{eqnarray}
T_1[n] = \sum_{k=1}^n c^k \left(\frac{1-q}{1-q^k}\right)^2,~~~c \in \mathbb{C} \backslash \{0\}. 
\end{eqnarray}
%-----------------------------------------------------------------------------------------------------
It obeys the recurrence
%-----------------------------------------------------------------------------------------------------
\begin{eqnarray}
&& c \left(q^{n+1}-1\right)^2 T_1[n-1]+
   \left(2 c q^{n+1}-c q^{2 (n+1)}-c+2
   q^{n+2}
-q^{2(n+1)+2}-1\right) T_1[n]
\nonumber\\ &&
+\left(q^{n+2}-1\right)^2 T_1[n+1] = 0,~~n \geq 1, 
\end{eqnarray}
%-----------------------------------------------------------------------------------------------------
with
%-----------------------------------------------------------------------------------------------------
\begin{eqnarray}
T_1[0] = c,~~T_1[1] = c + \frac{c^2
   (1-q)^2}{\left(1-q^2\right)^2}. 
\end{eqnarray}
%-----------------------------------------------------------------------------------------------------

The recurrence for $f[n;q]$ is obtained by
%-----------------------------------------------------------------------------------------------------
\begin{eqnarray}
&& 
     2 \left(q^n-1\right)^3 f[n;q]
   + \left(q^{n+1}-1\right) \left(q^n \left((2 (q-2) q+1) q^n+2\right)-1\right)
     f[n+1;q]
\nonumber\\ &&
   + \left(q^{n+1}-1\right) \left(1-q^{n+2}\right)\left((q (4 q-1)+2)q^n-5\right)
     f[n+2;q]
\nonumber\\ &&
   + 2 \left(q^{n+1}-1\right)\left(1-q^{n+2}\right) \left(q^{n+3}-1\right) f[n+3;q]
   = 0,
\end{eqnarray} 
%----------------------------------------------------------------------------------------------------- 
with 
%----------------------------------------------------------------------------------------------------- 
\begin{eqnarray} 
&& f[0;q] = f[1;q] = f[2;q] = 0, f[3;q] = \frac{1}{1 + 2 q + 2 q^2 + q^3}. 
\end{eqnarray} 
%----------------------------------------------------------------------------------------------------- 

The related $q$-shift relation reads 
%-----------------------------------------------------------------------------------------------------
\begin{eqnarray} 
&& q^6 (x-2) (x+1) (2 x+1) F(x)
  -q^3 (q (q (q (x (x+1) (8 x-9)-2)-x (3 x+4)-2)-7 x-2)
\nonumber\\ &&
-2) F(q x)
  +q \left(12 q^5 x^3+(q (2 q-7)-1) q^3 x^2-2 \left(q^2+q+1\right) (2 q (q+1)+1) q
   x
\right.
\nonumber\\ && \left. 
-2 \left(q \left(q^2+q+2\right)+1\right) 
q-2\right) F\left(q^2 x\right)
  +\left(-8 q^6 x^3+(q (5-2 (q-1) q)-1) q^3 x^2
\right.
\nonumber\\ && \left.
+\left((q (4 q+3)+10) q^2+q+2\right) q x+2
   \left(q^2+q+1\right) q+2\right) F\left(q^3 x\right)
  +(q x+1) \left(q^2 x-2\right) 
\nonumber\\ && 
\left(2 q^3 x+1\right) F\left(q^4 x\right) = 0,
\end{eqnarray}
%-----------------------------------------------------------------------------------------------------
with
%-----------------------------------------------------------------------------------------------------
\begin{eqnarray}
\langle 1\rangle [F(x)] = 0,~~\langle
   x\rangle [F(x)] = 0,~~
~~\left\langle
   x^2\right\rangle [F(x)] = 0,~~ 
~~\left\langle
   x^3\right\rangle [F(x)] 
= \frac{1}{(q+1)
   \left(q^2+q+1\right)}.
\end{eqnarray}
%-----------------------------------------------------------------------------------------------------
The $q$-differential equation reads
%-----------------------------------------------------------------------------------------------------
\begin{eqnarray}
&&
\biggl[(q x+1) \left(q^2 x-2\right) \left(2 q^3 x+1\right) ({\cal D}_x^q)^4
+(q (q (x (2 q ((q (q+2)+3) x+q)-3)-4)-2 x-3)
\nonumber\\ &&
-5) ({\cal D}_x^q)^3
   +(2 q((q (q+3)+3) x+q)-1) ({\cal D}_x^q)^2
   +2 {\cal D}_x^q\biggr] F[x;q] = 0,
\end{eqnarray}
%-----------------------------------------------------------------------------------------------------
where 
%-----------------------------------------------------------------------------------------------------
\begin{eqnarray}
F[0] = F^{(1)}[0] = F^{(2)}[0] = 0.
\end{eqnarray}
%-----------------------------------------------------------------------------------------------------

The expansion coefficient of the derivative of $\HA[-1/2, 2, -1, x]$, $2 
\HA[2,-1,x]/(1+2x)$, is
%-----------------------------------------------------------------------------------------------------
\begin{eqnarray}
f[n] = (-2)^{1+n}\left[
S_{1,1}\left({{-\frac{1}{4},-2},n}\right) -S_2\left({{\frac{1}{2}},n}\right) \right],
\end{eqnarray}
%-----------------------------------------------------------------------------------------------------
resulting into
%---------------------------------------------------------------------------------------------$
\begin{eqnarray}
\label{GHS4}
f^{(1)}[n;q] = (-2)^{1+n} \left[ 
\sum_{k=1}^n 
\left(-\frac{1}{4}\right)^k 
\frac{(1-q)}{(1-q^k)} \sum_{l=1}^k (-2)^l \frac{(1-q)}{(1-q^l)} - 
\sum_{k=1}^n  \left(\frac{1}{2}\right)^k \left(\frac{1-q}{1-q^k}\right)^2
\right].
\end{eqnarray}
%---------------------------------------------------------------------------------------------$
The hierarchy equation
follows from Eqs.~(\ref{eq:diffq}, \ref{eq:GHS2}) and (\ref{GHS4}) by 
%---------------------------------------------------------------------------------------------$
\begin{eqnarray}
{\cal D}_x^q \sum_{n=0}^\infty f[n;q] x^n = \sum_{n=1}^\infty f[n;q] \frac{1-q^n}{1-q} 
x^{n-1}. 
\end{eqnarray}
%---------------------------------------------------------------------------------------------$

%-----------------------------------------------------------------------------------------------------
\subsection{The \boldmath $q$-deformed cyclotomic harmonic polylogarithms}
\label{sec:45}
%-----------------------------------------------------------------------------------------------------

\vspace*{1mm}
\noindent
We consider the following cyclotomic harmonic polylogarithm, 
cf.~Ref.~\cite{Ablinger:2011te},
%--------------------------------------------------------------------------------------------
\begin{eqnarray}
\text{G}\left[\left\{\frac{1}{1-\tau+\tau^2},\frac{1}{1+\tau}\right\},x\right]
\end{eqnarray}
%--------------------------------------------------------------------------------------------
as an instructive example and construct its $q$-extension.
The associated function $f[n]$ is obtained by
%--------------------------------------------------------------------------------------------
\begin{eqnarray}
f[n] &=&
-\frac{(-1)^{2/3} \big(
        -(-1)^{2/3}\big)^n}{\big(1+\sqrt[3]{-1}\big) n}
S_1\big({{-\sqrt[3]{-1}},n}\big)
+\frac{(-1)^{2/3 +n/3}}{\big(1+\sqrt[3]{-1}\big) n}
S_1\big({{(-1)^{2/3}},n}\big),
\end{eqnarray}
%--------------------------------------------------------------------------------------------
cf.~Ref.~\cite{Blumlein:2026clg}, with
%--------------------------------------------------------------------------------------------
\begin{eqnarray}
\label{eqCY1}
f[n;q] &=& \frac{a^2}{\big(1+a\big) \{n\}_q} \Biggl[
-(-a^2)^n
S_1\big({{-a},n;q}\big)
+ a^{n}
S_1\big(a^2,n;q)\Biggr],~~a \equiv (-1)^{1/3}.
\end{eqnarray}
%--------------------------------------------------------------------------------------------
The recurrence reads
%--------------------------------------------------------------------------------------------
\begin{eqnarray}
&& -\left(q^n-1\right) \left(q^{n+1}-1\right) f[n;q]
-\left((q-1) \left(q^{n+1}-1\right) q^{n+1}\right) f[n+1;q]
\nonumber\\ &&
+(q-1) 
   \left(q^{n+2}-1\right)
   q^{n+1} f[n+2;q]
-
   \left(q^{n+2}-1\right)
   \left(q^{n+3}-1\right) f[n+3;q] = 0,
\end{eqnarray}
%--------------------------------------------------------------------------------------------
with 
%--------------------------------------------------------------------------------------------
\begin{eqnarray}
f[0;q] = f[1;q] = 0.
\end{eqnarray}
%--------------------------------------------------------------------------------------------
%--------------------------------------------------------------------------------------------
The $q$-shift relation is
%--------------------------------------------------------------------------------------------
\begin{eqnarray}
&& 
 (1 - q)^2 q x^2 
- q \left(x^3+1\right) F[x]
+ (x+1) \left(q (q+1)
   x^2-2 q x+q+1\right) F[q x]
\nonumber\\ &&
   -(x+1) (q x (q x-1)+1)
   F\left[q^2 x\right] = 0,
\end{eqnarray}
%--------------------------------------------------------------------------------------------
with
%--------------------------------------------------------------------------------------------
\begin{eqnarray}
   \langle 1\rangle
   [F(x)] = \langle x\rangle
   [F(x)]=0.
\end{eqnarray}
%--------------------------------------------------------------------------------------------
The $q$-differential equation is
%--------------------------------------------------------------------------------------------
\begin{eqnarray}
\biggl[(q x (q x-1)+1) ({\cal D}_x^{q})^2
- (x+1) (q
   x+x-1) {\cal D}_x^{q}\biggr] F[x;q]
= \frac{1}{1+x},
\end{eqnarray}
%--------------------------------------------------------------------------------------------
with 
%--------------------------------------------------------------------------------------------
\begin{eqnarray}
F[0;q] = {\cal D}_x^{q} [F][0;q] = 0.
\end{eqnarray}
%--------------------------------------------------------------------------------------------
By applying the $q$-differential operator one can check the hierarchy also in the
present case. One has
%--------------------------------------------------------------------------------------------
\begin{eqnarray}
\label{eq:hier44b}
f^{(1)}[n,q] = -\frac{1}{1+a} \big[
        a \big(
                -a^2\big)^n S_1({{-a},n})
        +a^n S_1\big({{a^2},n}\big)
\big]. 
\end{eqnarray}
%--------------------------------------------------------------------------------------------
From this it follows 
%---------------------------------------------------------------------------------------------$
\begin{eqnarray}
{\cal D}_x^q \sum_{n=0}^\infty f[n;q] x^n = \sum_{n=1}^\infty f[n;q] \{n\}_q
x^{n-1},
\end{eqnarray}
%---------------------------------------------------------------------------------------------$
cf.~Eqs.~(\ref{eqCY1}) and  (\ref{eq:hier44b}).

%-----------------------------------------------------------------------------------------------------
\subsection{The \boldmath $q$-deformed  harmonic polylogarithms inspired by 
quadratic forms}
\label{sec:46}
%-----------------------------------------------------------------------------------------------------

\vspace*{1mm}
\noindent 
An example of this class of iterated integrals is, cf.~Ref.~\cite{Ablinger:2021fnc}, 
%-----------------------------------------------------------------------------------------------------
\begin{eqnarray}
T_3 = 
\lefteqn{\int_0^x dy_1 \frac{1}{y_1^2+y_1-2}
\int_0^{y_1} dy_2 \frac{1}{y_2 + 1}
\int_0^{y_2} dy_3 \frac{1}{y_3 - 1} }
\nonumber\\
&=&
\sum_{n=1}^\infty
\frac{1}{3 n} \Biggl[
        \frac{1}{2} S_2({n})
        -\frac{1}{2} S_{-1}^2({n})
        -\left(
                -\frac{1}{2}\right)^n S_2({{-2},n})
        +\left(
                -\frac{1}{2}\right)^n S_{1,1}({{2,-1},n}) \Biggr] x^n. 
\nonumber\\
\end{eqnarray}
%-----------------------------------------------------------------------------------------------------
The $q$-extension of the central function $f[n]$ of $T_3$ reads
%-----------------------------------------------------------------------------------------------------
\begin{eqnarray}
f[n;q] &=& \frac{1}{3} \frac{1-q}{1-q^n} 
\Biggl\{
\frac{1}{2} 
\sum_{k=1}^n \left(\frac{1-q}{1-q^k}\right)^2
- \frac{1}{2} 
\left[\sum_{k=1}^n (-1)^k \frac{1-q}{1-q^k} \right]^2
-\left(-\frac{1}{2}\right)^n \sum_{k=1}^n (-2)^k \frac{1-q}{1-q^k}
\nonumber\\ &&
+\left(-\frac{1}{2}\right)^n \sum_{k=1}^n 2^k \frac{1-q}{1-q^k} \sum_{l=1}^k (-1)^l
\frac{1-q}{1-q^l}
\Biggr\}
\end{eqnarray}
%-----------------------------------------------------------------------------------------------------
with
%-----------------------------------------------------------------------------------------------------
\begin{eqnarray}
\{T_3\}_q = \sum_{n=0}^\infty f[n;q] x^n.
\end{eqnarray}
%-----------------------------------------------------------------------------------------------------

The recursion for $f[n,q]$ is of 5th order and has the following structure
%-----------------------------------------------------------------------------------------------------
\begin{eqnarray}
\label{eq:s46:larg}
&&  q^2 (q^n-1) ((q^{1 + n}-1)^2) (q^{2 + n}-1) 
[p_1(q) 
+ q^n p_2(q) 
+ q^{2n} p_3(q)] f[n;q]  
\nonumber\\ &&
 +(q-1) (q^{1 + n}-1) 
(q^{2 + n}-1)
[p_4(q) 
+ q^n p_5(q) 
+ q^{2n} p_6(q)
+ q^{3n} p_7(q)
+ q^{4n} p_8(q)] f[n+1;q]  
\nonumber\\ &&
+ (q^{2 + n}-1) 
[p_9(n) 
+ q^n p_{10}(q) 
+ q^{2n} p_{11}(q)
+ q^{3n} p_{12}(q)
+ q^{4n} p_{13}(q)
+ q^{5n} p_{14}(q)] f[n+2;q]  
\nonumber\\ &&
+ (q^{3 + n}-1)
[p_{15}(q) 
+ q^n p_{16}(q) 
+ q^{2n} p_{17}(q)
+ q^{3n} p_{18}(q)
+ q^{4n} p_{19}(q)
+ q^{5n} p_{20}(q)] f[n+3;q]  
\nonumber\\ &&
+ ( q^{4 + n}-1)^2
[p_{21}(q) 
+ q^n p_{22}(q) 
+ q^{2n} p_{23}(q)
+ q^{3n} p_{24}(q)
+ q^{4n} p_{25}(q)] f[n+4;q]  
\nonumber\\ &&
+ 2 (q^{4 + n}-1) (q^{5 + n}-1)^3
[p_{26}(q)
+ q^n p_{27}(q)
+ q^{2n} p_{28}(q)
] f[n+5;q] = 0,
\end{eqnarray} 
%-----------------------------------------------------------------------------------------------------
with large polynomials $p_i(q)$.\footnote{The recurrence is given in 
computer-readable form as ancillary file to this paper.}
Analogously to previous cases, one can derive also the $q$-differential equation and 
$q$-shift relation, which also form very large expressions.
%-----------------------------------------------------------------------------------------------------
The hierarchy relation is 
%-----------------------------------------------------------------------------------------------------
\begin{eqnarray}
{\cal D}_x^{q}
\{T_3\}_q = \sum_{n=1}^\infty f[n;q] \frac{1-q^n}{1-q} x^{n-1},
\end{eqnarray}  
%-----------------------------------------------------------------------------------------------------
which is equal to the $q$-extension of the moments of $dT_3/dx 
= -\HA_{-1,1}(x)/(x^2+x-2)$.
%-----------------------------------------------------------------------------------------------------
The analysis of the recurrence Eq.~(\ref{eq:s46:larg}) by using {\tt Sigma} shows, that 
the 
minimal recurrence is actually of order {\sf o = 4}. The larger recurrence was obtained
by joining recurrences for individual summands.

%-----------------------------------------------------------------------------------------------------
\subsection{The \boldmath $q$-deformed iterated integrals containing square-root 
valued letters}
\label{sec:47}
%-----------------------------------------------------------------------------------------------------

\vspace*{1mm}
\noindent 
Square-root valued iterated integrals have been considered in 
Refs.~\cite{Ablinger:2014bra,Blumlein:2023vwi}. Their Mellin transforms lead to 
nested 
sums with central binomials at different places in the numerator or denominator.
They arise in higher order 
perturbative calculations. In particular we refer to the letters (3.25--3.63) of the 
alphabet given in Ref.~\cite{Ablinger:2014bra}. This class of functions is the most 
involved one found in quantum 
field theoretical calculations, which still obey first order factorizing differential 
and difference equations.

As a typical example we consider the $q$-extension of the function 
%-----------------------------------------------------------------------------------------------------
\begin{eqnarray}
T_4(x) = \sum_{n=1}^\infty \frac{4^n}{\displaystyle n^2 \binom{2 n}{n}} S_1(n) x^n,  
\end{eqnarray}
%-----------------------------------------------------------------------------------------------------
with
%-----------------------------------------------------------------------------------------------------
\begin{eqnarray}
\label{T4xq}
T_4(x;q) = \sum_{n=1}^\infty 4^n \left(\frac{1-q}{1-q^n}\right)^2 
\frac{\displaystyle \prod_{k=1}^n 
(1-q^k)^2}{\displaystyle \prod_{k=1}^{2n} 
(1-q^k)} 
\sum_{k=1}^n \frac{(1-q)}{(1-q^k)} 
~x^n = \sum_{n=1}^\infty f[n;q] x^n.
\end{eqnarray}
%-----------------------------------------------------------------------------------------------------
We obtain the following recurrence for the function $f[n;q]$ 
%-----------------------------------------------------------------------------------------------------
\begin{eqnarray}
&& 
16
   \left(q^n-1\right)^2
   \left(q^{n+1}-1\right)^2
   f[n;q] 
-4 \left(q^{n+1}+1\right)
   \left(q^{2 n+1}-1\right)
   \left((q+1) q^{n+1}-2\right)
   \left(q^{n+1}-1\right)^2
\nonumber\\ && \times
   f[n+1;q]
+\left(q^{n+1}+1\right
   ) \left(q^{n+2}-1\right)^2
   \left(q^{n+2}+1\right)
   \left(q^{2 n+1}-1\right)
   \left(q^{2 n+3}-1\right)
   f[n+2;q]
= 0, \nonumber\\ \end{eqnarray} 
%----------------------------------------------------------------------------------------------------- 
with 
%----------------------------------------------------------------------------------------------------- 
\begin{eqnarray} 
f[0;q] = 0,~~ f[1;q] = \frac{4}{1+q}. 
\end{eqnarray} 
%----------------------------------------------------------------------------------------------------- 
In this sense the case of square-root iterated integrals is not different from the other 
iterated integrals obeying first order factorizing differential equations. 

The $q$-shift 
relation is given by 
%----------------------------------------------------------------------------------------------------- 
\begin{eqnarray} && 
\label{eq:47a}
q^5 (1-4x)^2 F[x]
+ Q_1 F[q x]
+ Q_2 F[q^2 x]
+ Q_3 F[q^3 x]
+ Q_4 F[q^4 x]
+ Q_5 F[q^5 x]
\nonumber\\ &&
+ Q_6 F[q^6 x]
+ (q-1) F[q^7 x]
+ F[q^8 x]
+ 8 (q-1)^4 q^4 x = 0,
\end{eqnarray}
%-----------------------------------------------------------------------------------------------------
with the polynomials
%-----------------------------------------------------------------------------------------------------
\begin{eqnarray}
Q_1 &=& -q^4 (q (8 x-1) (4 (q+1) x-1)-1), \nonumber\\ 
Q_2 &=& q^2 \left(q^2 (q (4 x (4 (q (q+4)+1) x-q+1)-1)+8 x-2)-1\right), \nonumber\\ 
Q_3 &=& 	-q
   \left(q \left(q (q+1)
   \left(q \left(4 x \left(8
   q^2
   x+q+3\right)-1\right)+1\right)-1\right)+1\right), \nonumber\\
Q_4 &=& q
   \left(q^2 \left(4 q x
   \left(q \left(4 q^2
   x+q+2\right)-1\right)+2
   q+1\right)+q+2\right), \nonumber\\
Q_5 &=&
\left(q^2 (q (4
   (q+3) q x-q+1)-1)+1\right), \nonumber \\
Q_6 &=&
-\left(4
   (q+1) q^4 x+q^3+2 q+1\right). 
\end{eqnarray}
%-----------------------------------------------------------------------------------------------------
The initial values for Eq.~(\ref{eq:47a}) are $\langle x^k \rangle [F[x]] = f[k;q], k \in 
[0,6]$ given by the first 
expansion coefficients
of $T_4(x;q)$, Eq.~(\ref{T4xq}).

The $q$-differential equation is obtained by 
%-----------------------------------------------------------------------------------------------------
\begin{eqnarray}
&& \biggl\{
8 + Q_7 {\cal D}_x^{q} 
+Q_8 ({\cal D}_x^{q})^2
+Q_9 ({\cal D}_x^{q})^3
+Q_{10} ({\cal D}_x^{q})^4
+Q_{11} ({\cal D}_x^{q})^5
+Q_{12} ({\cal D}_x^{q})^6
+Q_{13} ({\cal D}_x^{q})^7
\nonumber\\ &&
+(q-1)^4 q^{24} x^7 ({\cal D}_x^{q})^8 \biggr\} F[x;q] = 0, 
\end{eqnarray}
%-----------------------------------------------------------------------------------------------------
with the polynomials 
%-----------------------------------------------------------------------------------------------------
\begin{eqnarray}
Q_7 &=& 2 (q+1) \left(2 q^2 x (4 x-1)+4 q x (2 x-1)-1\right), 
\nonumber\\
Q_8 &=& -(q+1) x \left(4 q^7 x+q^6 (12 x-1)-2 q^5 \left(8 x^2-12 x+1\right)+q^4 
\left(-32 x^2+28 x-3\right) \right.
\nonumber\\ && \left.
+q^3 \left(-48 x^2+28 x-3\right)-2 q^2 
\left(8 x^2-6 x+1\right)-q-2\right), 
\nonumber\\
Q_9 &=&
q x^2 (q^{12}+q^{11} (2-4 x)+q^{10} (5-12 x)+q^9 (7-24 x)+q^8 (11-40 x)
\nonumber\\ && 
+4 
q^7 \left(4 x^2-12 x+3\right)+q^6 \left(32 x^2-48 x+13\right)+q^5 \left(48 x^2-28 
x+9\right)
\nonumber\\ &&
+q^4 \left(32 x^2-8 x+6\right)+q^3 (12 x+1)+8 q^2 x-q-2),
\nonumber\\
Q_{10} &=&
q^3 x^3 (q^{15}+q^{14}+2 q^{13}-4 q^{12} (x-1)+q^{11} (5-8 x)+q^{10} (5-12 x)
\nonumber\\ && 
+q^9 
(7-16 x)+q^8 (4-16 x)+q^7 (2-4 x)+16 q^6 x^2+q^5 (8 x-3)+2 q^4 (8 x-3)
\nonumber\\ &&
+q^3 (4 x-2)-3 q^2 -2 q+1),
\nonumber\\
Q_{11} &=&
(q-1) q^7 x^4 (q^{14}+q^{13}+2 q^{12}+3 q^{11}+5 q^{10}+q^9 (5-4 x)+q^8 (7-8 x)
\nonumber\\ &&
+q^7 
(5-8 x)+q^6 (6-8 x)+q^5 (3-8 x)+q^4 (1-4 x)+q^3 (8 x-2)-3 q-2), 
\nonumber\\
Q_{12} &=& (q-1)^2 q^{11} x^5 (q^{12}+q^{11}+2 q^{10}+2 q^9+3 q^8+4 q^7
\nonumber\\ &&
+4 q^6+q^5 
(3-4 x)
+q^4 (3-4 x)+q^3+2 q^2-q-1), 
\nonumber\\
Q_{13} &=& 
(q-1)^3 \left(q^6+q^5+q^4+q^3+q^2+q+2\right) q^{18} x^6,
\end{eqnarray}
%-----------------------------------------------------------------------------------------------------
where 
$({\cal D}_x^{q})^{k} F[0;q] = \{n!\}_q f[k;q],~~~k \in [1,7]$.

Given the complexity of the expression (\ref{T4xq}) compared to other cases discussed 
before, the recurrence is small. However, much larger 
$q$-shift and $q$-differential equations are obtained.
In the same way as for (\ref{T4xq}) one may solve more involved cases, leading to larger
determining equations.

With this we have found for all presently known first order factorizing solutions of
special functions emerging in higher order calculations of single scale Feynman diagrams
the respective $q$-extensions, either in general form or by algorithmic steps to be 
applied to the respective cases. They are uniquely defined by their $q$-recursion 
relations, their $q$-differential equations, and $q$-shift relations.
%----------------------------------------------------------------------------------------------------------------
\section{Shuffle and quasi-shuffle algebras}  
\label{sec:5}
%----------------------------------------------------------------------------------------------------------------
	
\vspace*{1mm}
\noindent
Also in the case of $q$-deformed iterated integrals and nested sums the products of 
the respective quantities can be described by shuffle and quasi-shuffle relation, 
cf.~Refs.~\cite{Blumlein:1998if,Hoffman:99,Moch:2001zr,Blumlein:2003gb}, since these
are purely index-based and induced by the respective alphabets.

The products of two iterated integrals $G_{a_1,...,a_l}(x)$ and $G_{b_1,...,b_m}(x)$ is the 
sum of all iterated integrals of weight $l+m$ over all shuffles  
%-----------------------------------------------------------------------------------------------------
\begin{eqnarray}
\{\{c_1,...,c_{l+m}\}\} = \{a_1,...,a_l\} \shuffle \{b_1,...,b_m\}, 
\end{eqnarray}
%-----------------------------------------------------------------------------------------------------
where the shuffle product is denoted by $\shuffle$. The indices $a_i, b_j$ refer to 
the alphabet $\mathfrak{A}_F$ of functions to be iterated, where the order of 
letters of the set $\{a_1,...,a_l\}$ and $\{b_1,...,b_m\}$ is preserved, while 
all other combinations are allowed,
%-----------------------------------------------------------------------------------------------------
\begin{eqnarray}
\label{eq:shuf1}
G_{a_1,...,a_l}(x) \cdot G_{b_1,...,b_m}(x) = \sum_{C \in \{\{c_1,...,c_{l+m}\}\}} G_C(x).
\end{eqnarray}
%-----------------------------------------------------------------------------------------------------
The nested sums obey quasi-shuffle products, since there are additional trace 
terms, cf. Refs.~\cite{Moch:2001zr,Blumlein:2003gb}, which apply also to their 
$q$-extension. 
We illustrate the principal structure for 
the $q$-extended generalized harmonic sums. For the product of $q$-extended generalized
harmonic sums,
$S_{a_1,...,a_l}(c_1,...,c_l;N;q)$ and $S_{b_1,...,b_m}(d_1,...,d_m;N;q)$, one 
obtains
%-----------------------------------------------------------------------------------------------------
\begin{eqnarray}
\label{eq:SHgHS}
\lefteqn{
S_{a_1,...,a_l}(c_1,...,c_l;N;q) \cdot S_{b_1,...,b_m}(d_1,...,d_m;N;q) =} 
\nonumber\\ && 
\hspace*{4cm} \sum_{n=1}^N \frac{c_1^n}{\{n\}_q^{a_1}} S_{a_2,...,a_l}(c_2,...,c_l;N;q) 
\cdot 
S_{b_1,...,b_m}(d_1,...,d_m;N;q)
\nonumber\\ && \hspace*{4cm}
+ \sum_{n=1}^N \frac{d_1^n}{\{n\}^{b_1}_q} S_{a_1,...,a_l}(c_1,...,c_l;n;q) \cdot 
S_{b_2,...,b_m}(d_2,...,d_m;n;q) 
\nonumber\\ && \hspace*{4cm}
- \sum_{n=1}^N \frac{(c_1 \cdot d_1)^n}{\{n\}^{a_1+b_1}_q} 
S_{a_2,...,a_l}(c_2,...,c_l;n;q) \cdot 
S_{b_2,...,b_m}(d_2,...,d_m;n;q), 
\nonumber\\ 
\end{eqnarray}
%-----------------------------------------------------------------------------------------------------
where additional terms beyond the shuffle-contributions emerge. 

Various explicit examples in the case of 
harmonic sums are given in Refs.~\cite{Blumlein:1998if,Blumlein:2003gb}. One example is
%-----------------------------------------------------------------------------------------------------
\begin{eqnarray}
S_b(N;q) \cdot S_{a_1,...,a_l}(N;q) &=& S_b(N;q) \shuffle S_{a_1,...,a_l}(N;q)
\nonumber\\ & &
- S_{b \wedge a_1,a_2,...,a_l}(N;q) - ... - S_{a_1,a_2,...,b \wedge a_l}(N;q)
\end{eqnarray}
%-----------------------------------------------------------------------------------------------------
with
%-----------------------------------------------------------------------------------------------------
\begin{eqnarray}
S_b(N;q) \shuffle S_{a_1,...,a_l}(N;q) =
S_{b,a_1,...,a_l}(N;q) + 
S_{a_1,b,a_2,...,a_l}(N;q) + ... + 
S_{a_1,...,a_l,b}(N;q), 
\end{eqnarray}
%-----------------------------------------------------------------------------------------------------
and
%-----------------------------------------------------------------------------------------------------
\begin{eqnarray}
a \wedge b = (|a|+|b|) {\rm sign}(a) {\rm sign}(b). 
\end{eqnarray}
%-----------------------------------------------------------------------------------------------------
For the counting of independent iterative integrals and sums, 
see Refs.~\cite{WITT1,WITT2,LYNDON1,LYNDON2,RADFORD} and
Ref.~\cite{Blumlein:2003gb}.

Let us consider the relations, cf.~Ref.~\cite{Blumlein:2003gb},
%-----------------------------------------------------------------------------------------------------
\begin{eqnarray}
\label{eq:SH1}
S_{1,-1} &=&  - S_{-1,1} + S_{1} S_{-1} + S_{-2} 
\\
\label{eq:SH2}
S_{1,-1,-2} &=&
S_{-2} S_{1,-1} + S_{1,3} + S_{3,1} - S_{-2,1,-1} - S_{-1,-2,1} - S_1 S_{-2,-1}
+ S_{-1} S_{-2,1}, 
\\
S_{-1,2,1,-2} &=& S_{-3,1,-2} + S_{2,-2,-2} + S_{2,1,3} 
- S_{2,-1,1,-2} - S_{2,1,-2,-1} - S_{2,1,-1,-2} + 
S_{-1} S_{2, 1, -2}, 
\nonumber\\
\\
\label{eq:SH4}
S_{1,2,3,4} &=& S_{2,3,5} + S_{2,4,4} + S_{3,3,4} - S_{2,1,3,4} 
- S_{2,3,1,4} - S_{2,3,4,1} + S_1 S_{2, 3, 4}, 
\end{eqnarray}
%-----------------------------------------------------------------------------------------------------
where we applied the short-hand notation $S_{\vec{a}}(N;q) \equiv S_{\vec{a}}$.
The index pattern implies that the double sum of Eq.~(\ref{eq:SH1} needs 
one defining double sum
at the r.h.s. For the triple sum Eq.~(\ref{eq:SH2}, two triple sums are needed. 
For the present pattern in the case of the quadruple sums we used other 
four-letter relations to shorten the given expression, cf.~\cite{Blumlein:2003gb}.

We first verify the validity of the shuffle-relations for $q$-extended harmonic sums at fixed values of 
$N$ up to a maximal value $N_{\rm max} = 20$, expanding the $q$-rational functions. 
Next we calculate the $q$-recurrences of both sides of Eqs.~(\ref{eq:SH1}--\ref{eq:SH4}) 
and show 
that they are identical. 
We obtain the following recurrences for $S_{1,-1}(N;q)$ and 
for $S_{1,-1,-2}(N;q)$
%-----------------------------------------------------------------------------------------------------
\begin{eqnarray}
&& f[n;q]
   \left(q^{n+2}-1\right)
   \left(q^{n+3}-1\right)
+f[n+1;q]
   \left(q^{n+3}-1\right)
   \left(-q^{n+2}-q^{n+3}+q^{n+4}
   +1\right)
\nonumber\\ &&
-f[n+2;q]
   \left(q^{n+3}-3 q^{n+4}-q^{2
   n+6}+q^{2 n+7}+q^{2
   n+8}+1\right)
+ f[n+3;q]
   \left(q^{n+4}-1\right)^2 = 0,
\nonumber\\
\\
&& 
 f[n;q]
   \left(q^{n+4}-1\right)^2
   \left(-q^{n+2}-q^{n+3}+q^{2
   n+5}+1\right)
+ f[n+1;q]
   \left(q^{n+2}+3 q^{n+3}+2
\right.
\nonumber\\ && \left.
   q^{n+4}+q^{3 (n+4)}+2 
   q^{n+5}-q^{2 (n+5)}-q^{2
   n+5}-3 q^{2 n+6}-4 q^{2
   n+7}-q^{2 n+8}-2 q^{2
   n+9}+2 q^{3 n+9} \right.
\nonumber\\ && \left.
+3 q^{3
   n+10}+q^{3 n+13}+q^{3
   n+14}-q^{4 n+13}-q^{4
   n+14}+q^{4 n+15}-q^{4
   n+17}-2\right)
\nonumber\\ &&
- f[n+2;q]
\left((q-1) 
   \left(q^{n+3}+3 q^{n+4}-2
   q^{n+5}-q^{2 (n+5)}+q^{3
   (n+5)}+q^{n+6}-2 q^{2
   n+7}-q^{2 n+8} \right. \right.
\nonumber\\ && \left. \left.
+2 q^{2
   n+9}-q^{2 n+11}+q^{3
   n+11}-q^{3 n+13}+q-2\right)
   q^{n+3}\right)
+ f[n+3;q]
   \left(-3 q^{n+4}-4 q^{3
   (n+4)}
\right.
\nonumber\\ && \left.
+q^{4 (n+4)} 
-5
   q^{n+5}+8 q^{2 (n+5)}-5
   q^{3 (n+5)}+q^{4 (n+5)}+5
   q^{2 n+8}-q^{2 n+9}+2 q^{3
   n+13}-q^{3 n+14} \right.
\nonumber\\ && \left.
-q^{4
   n+18}+q^{4
   n+19}+2\right)
- f[n+4;q]
   \left(q^{n+5}-1\right)^4 = 0.
\end{eqnarray}
%-----------------------------------------------------------------------------------------------------
Similar, but much larger, recurrences are obtained for $S_{-2,1,2,-1}(N;q)$ and 
$S_{1,2,3,4}(N;q)$
and their shuffle-product representations.

Complementary to this, we use the algorithms of the package {\tt Sigma}
\cite{SIG1,SIG2} 
to reduce the $q$-sums for general values of $N$ to show the identities. The latter two algorithms can be 
applied to each individual case of products of nested harmonic sums.
Let us illustrate this for Eq.~(\ref{eq:SH2}). One forms the $q$-sum tower 
%-----------------------------------------------------------------------------------------------------
\begin{eqnarray}
\label{eq:SH20}
\{S_1, 
S_{-1}, 
S_{-2},
S_{-2,-1},
S_{-2,1},
S_{1,-1}, 
S_{1,3}, 
S_{3,1},
S_{-2,1,-1}, 
S_{-1,-2,1}\},
\end{eqnarray}
%-----------------------------------------------------------------------------------------------------
over which other sums shall be expressed.
By using the {\tt Sigma}-command
%-----------------------------------------------------------------------------------------------------
\begin{eqnarray}
&&
{\tt SigmaReduce[qS[q,1,-1,-2, n] , n, Tower \rightarrow tower, SimpleSumRepresentation  
}\nonumber\\ &&
{\tt \rightarrow False, qCase \rightarrow q]
}
\end{eqnarray}
%-----------------------------------------------------------------------------------------------------
one obtains the r.h.s. of (\ref{eq:SH2}), the synonymous relation to the $q$-free case.
More generally, one can start with a tower containing all sums
needed to represent any nested sum at a given weight and cannot only prove but
even find any identity that is induced by a quasi-shuffle product.
The package {\tt Sigma} can be applied to discover the underlying identity.

Let us now turn to generalized harmonic sums in extending the sums in 
Eqs.~(\ref{eq:SH1}--\ref{eq:SH2})
by adding the numerator weights 
%-----------------------------------------------------------------------------------------------------
\begin{eqnarray}
\label{eq:SH10}
\left\{a, b, c \right\}
\end{eqnarray}
%-----------------------------------------------------------------------------------------------------
to the first and second sum and apply the same techniques as for the harmonic sums. We choose 
%-----------------------------------------------------------------------------------------------------
\begin{eqnarray}
\label{eq:SH10a}
\left\{a = 3,~~b = - \frac{1}{2},~~c = \frac{1}{5} \right\}
\end{eqnarray}
%-----------------------------------------------------------------------------------------------------
as concrete examples. The following recursions are obtained for the respective 
l.h.s, and r.h.s. of 
the $q$ extensions of $S_{1,1}(\{a,b\},n)$ and $S_{1,1,2}(\{a,b,c\},n)$,
%-----------------------------------------------------------------------------------------------------
\begin{eqnarray}
\label{eq:SH10b}
&& 9
   \left(q^{n+1}-1\right)
   \left(q^{n+2}-1\right)
   f[n;q]
+3
   \left(q^{n+2}-1\right)
   \left((q+1) (2 q-3)
   q^{n+1}+2\right)
   f[n+1,q]
\nonumber\\ &&
+\left(10
   q^{n+3}+(3-2 q (q+3)) q^{2
   n+4}-5\right) f[n+2;q]
+2 \left(q^{n+3}-1\right)^2
   f[n+3;q] = 0,
\end{eqnarray}
%-----------------------------------------------------------------------------------------------------
with
%-----------------------------------------------------------------------------------------------------
\begin{eqnarray}
f[0;q] = 0,~~f[1;q] = -\frac{3}{2},~~ 
f[2;q] = -\frac{3 (5 + 2 q (5 + q))}{4 (1 + q)^2},
\end{eqnarray}
%-----------------------------------------------------------------------------------------------------
and
%-----------------------------------------------------------------------------------------------------
\begin{eqnarray}
\label{eq:SH10c}
&& 
27
   \left(q^{n+1}-1\right)
   \left(q^{n+2}-1\right)
   \left(q^{n+3}-1\right)^2
   f[n;q]
+9
   \left(q^{n+2}-1\right)
   \left(q^{n+3}-1\right)
   \left(\left(\left(q
   \left(10 q^3+2
   q-1\right)
\right. \right. \right.
\nonumber\\ && \left. \left. \left.
-3\right) 
   q^{n+3}-20 q^3+q+3\right)
   q^{n+1}+8\right) f[n+1;q]
+3
   \left(q^{n+3}-1\right)
   \left(-5 (q (4 q+5)+6)
   q^{n+2} \right.
\nonumber\\ && \left.
+(2 q (2 q (18-5
   (q-1) q)+3)-3) q^{2 n+4}+(2
   q (q (5 q (q+1) (2
   q-3)-1)-3)+3) q^{3
   n+7}+25\right) 
\nonumber\\ && \times
f[n+2;q]
+2
   \left(q^{n+3} \left(-3 (5 q
   (9 q+2)-11) q^{n+3}+2 (5 q
   (q (7 q+6)-3)-6) q^{2
   n+6} \right. \right.
\nonumber\\ && \left. \left.
+\left(3-5 q^2 (2 q
   (q+3)-3)\right) q^{3
   n+9}+100
   q-12\right)-22\right)
   f[n+3;q]
\nonumber\\ &&
+20 \left(q^{n+4}-1\right)^4
   f[n+4;q] = 0,
\end{eqnarray}
%-----------------------------------------------------------------------------------------------------
with
%-----------------------------------------------------------------------------------------------------
\begin{eqnarray}
f[0;q] &=& 0,~~
f[1;q] = -\frac{3}{10},~~
f[2;q] = -\frac{3 (5 q (q (2 q (q+7)+27)+20)+22)}{100 (q+1)^4},~~
\nonumber\\ 
f[3;q] &=& - \frac{3 P_4(q)}{1000 (q+1)^4
   \left(q^2+q+1\right)^4},
\nonumber\\ 
 P_4(q) &=& 100 q^{12}+1100
   q^{11}+6050 q^{10}+20850
   q^9+49945 q^8+88240
   q^7+118510 q^6.
\end{eqnarray}
%-----------------------------------------------------------------------------------------------------

For the shuffle relations of the iterated integrals, Eq.~(\ref{eq:shuf1}), we consider
the formal series expansion around $x=0$ for the left and the right-hand sides, see 
Eq.~(\ref{eq:2}) 
or its extensions being modulated by a finite number  of powers in $\ln(x)$,
%-----------------------------------------------------------------------------------------------------
\begin{eqnarray}
\label{eq:SH11}
\sum_{n=0}^\infty f_l[n] x^n = \sum_{n=0}^\infty f_r[n] x^n.
\end{eqnarray}
%-----------------------------------------------------------------------------------------------------
Both $f_l[n]$ and $f_r[n]$ are sum-product structures, which can be $q$-extended as 
described before. The identity $f_l[n] = f_r[n]$ implies the identity
$f_l[n;q] = f_r[n;q]$ of the $q$-extensions. Therefore,
the $q$-extension preserves the shuffle product. Similarly as above, one may also use
the package {\tt Sigma} to discover the underlying identity.

Shuffle-algebras are Hopf-algebras \cite{HOPF,MILNER,SWEEDLER,Kreimer:1997dp}, 
with the product implied by the shuffle product, cf.~Ref.~\cite{REUTENAUER}.
%----------------------------------------------------------------------------------------------------------------
\section{Conclusions}  
\label{sec:6}
%----------------------------------------------------------------------------------------------------------------
	
\vspace*{1mm}
\noindent
Perturbative calculations in quantum field-theories are based on Feynman integrals, which 
describe all contributions up to a given power in the coupling constant. In the case of 
single or double-scale integrals, the respective integrals can be carried out analytically
and are given as functions in special function spaces either as iterated 
integrals or nested sums. 
This applies to the iterated integrals as
polylogarithms, Nielsen integrals, harmonic polylogarithms, generalized harmonic polylogarithms,
cyclotomic harmonic polylogarithms, iterated integrals implied by quadratic forms, and 
iterated integrals over square-root valued letters,
\cite{LEWIN1,LEWIN2,Devoto:1983tc,NIELSEN1,Kolbig:1983qt,Remiddi:1999ew,KUMMER1,KUMMER2,
KUMMER3,POINCARE1,LAPPO,CHEN,GONCHAROV,Moch:2001zr,Ablinger:2013cf,Ablinger:2011te,
Ablinger:2021fnc,Ablinger:2014bra}. For the nested sums we have considered the 
$q$-extensions of 
the harmonic sums, generalized harmonic 
sums, cyclotomic harmonic sums, nested sums implied by quadratic forms, and nested sums 
weighted by central binomial coefficients 
\cite{Vermaseren:1998uu,Blumlein:1998if,
Moch:2001zr,Ablinger:2013cf,Ablinger:2011te,Ablinger:2021fnc,Ablinger:2014bra}.
We considered the classes of 
iterative integrals obeying first 
order factorizing differential equations and nested sums obeying first order factorizing difference 
equations.

Like also the case in other special functions, 
cf.~Refs.~\cite{HEINE1,HEINE2,BAILEY,SLATER,EXTON,GASRHA,KOORNWINDER,PWZ,ANDREWS,KOEKOEK,
KACH,NIST,ISMAIL,KOEPF,ISMAIL1,JOHNSON}, 
one may construct the associated $q$-extension 
for these functions. From the quantum field theoretic point of view, the $q$-extension is implied
by the $q$-deformation of the commutation relation, Eq.~(\ref{eq:com}), and the associated 
dynamical equations.

The different spaces of the $q$-extended nested sums are determined by 
recursions of finite order and degree, depending on the deformation 
parameter $q$. Likewise, the spaces of iterated integrals are defined by $q$-differential and
$q$-shift relations of finite order and degree. 
For the single sums, the polylogarithms and Nielsen 
integrals, closed 
form $q$-extensions have been derived. For the higher sum- and function spaces algorithms 
exist
to derive the $q$-extension in the respective individual cases. In the present approach
the 
starting point to derive
the $q$-extensions is the Taylor expansion of the non $q$-deformed functions at $x = 0$, 
cf.~Eq.~(\ref{eq:2}). For the central functions $f[n;q]$ a $q$-recursion can be obtained
by guessing methods. One may also consider $q$-extensions by expanding around another
value as $x = x_0$ and find overlapping representations to map out the considered
region of $x$, beyond the convergence radius of the $q$-expansion around $x=0$.
The package {\tt Sigma} \cite{SIG1,SIG2} enables one to solve all derived $q$-recursions
for the functions $f[n;q]$ to obtain the respective initial representation. We used
this as an essential check of these relations.

Unlike the case for the $\mu$-deformation, Ref.~\cite{Blumlein:2026clg}, in the 
case of iterated integrals based on square-root valued letters, their $q$-extension 
remains in the same function class.
Both the $q$-extended nested sums and iterated integrals form (quasi-)shuffle algebras 
and therefore by the shuffle product Hopf algebras, cf.~Ref.~\cite{REUTENAUER}.
%----------------------------------------------------------------------------------------------------------------

\vspace*{5mm}
\noindent
{\bf Acknowledgment.}~~We thank Kolleg Mathematik Physik Berlin for financial support 
and C.~Koutschan, U.~K\"uhn, P.~Marquard, P.~Paule, K.~Sch\"onwald, and A.~Uncu for 
discussions. The research of A.M.G.was also funded by the National Academy of Sciences 
of Ukraine by its priority project No.~0122U000888. The research of C.S. was funded in 
part by the Austrian Science Fund (FWF) Grants DOI 10.55776/P33530 and 
10.55776/PAT1332123. 
%-----------------------------------------------------------------------------------------------------

\end{document}